\documentclass[10pt,twocolumn,twoside]{IEEEtran}
\pdfoutput=1

\usepackage[cmex10]{amsmath}
\usepackage{amssymb}
\usepackage{amsfonts}
\usepackage{mathrsfs}
\usepackage{comment}
\usepackage{cite}
\usepackage{graphicx}
\usepackage{rotating}
\usepackage{subfigure}
\usepackage{bm}
\usepackage{url}
\usepackage{standalone}
\usepackage{blindtext}
\usepackage{float}
\usepackage{hyperref}
\usepackage{bm}
\usepackage{makecell}
\usepackage{array}
\usepackage{subscript}
\usepackage{textcomp}
\usepackage{caption}

\newcolumntype{?}{!{\vrule width 2pt}}

\hypersetup{%
 colorlinks=true, %
 linkcolor=black,   
 anchorcolor=black, 
 citecolor=black,   
 filecolor=blue,   
 menucolor=blue,   
 runcolor=blue,    
 urlcolor=blue,    
}

\newcommand{\addstretch}[1]{\addtolength{#1}{\fill}}
\newenvironment{onepage}
  {\clearpage\flushbottom
   \addstretch{\baselineskip}
   \addstretch{\abovedisplayskip}
   \addstretch{\abovedisplayshortskip}
   \addstretch{\belowdisplayskip}
   \addstretch{\belowdisplayshortskip}
   \setlength{\parskip}{0pt}}
  {\clearpage}

\mathchardef\mhyphen="2D

\newcommand{\gr}[1]{\bm{#1}}

\newcommand{\q}{\gr{q}}

\newcommand{\Jn}{\mathrm{J}}

\newcommand{\nul}[1]{\mathcal{N}(#1)}

\newcommand{\range}[1]{\mathcal{R}(#1)}
\newcommand{\rangeperp}[1]{\mathcal{R}(#1)^\perp}

\newcommand{\n}{\gr{n}}

\newcommand{\x}{\gr{x}}
\renewcommand{\th}{\bm{\theta}}

\newcommand{\y}{\gr{y}}

\newcommand{\sgens}[1]{\mathbb{R}^{#1}}
\newcommand{\sgen}[2]{\mathbb{R}^{#1\times #2}}

\newcommand{\sm}{\mathbb{R}^m}
\newcommand{\sll}{\mathbb{R}^l}
\newcommand{\sk}{\mathbb{R}^k}

\newcommand{\lm}{\mathbb{R}^{l\times m}}

\newcommand{\nn}{\mathbb{R}^{n\times n}}

\newcommand{\ml}{\mathbb{R}^{m\times l}}
\newcommand{\mk}{\mathbb{R}^{m\times k}}

\newcommand{\mn}{\mathbb{R}^{m\times n}}

\newcommand{\tr}{\mathbb{R}^{t\times r}}

\newcommand{\me}[1]{\mathbb{E}[{#1}]}
\newcommand{\zt}{\textrm}
\newcommand{\xr}{\mathcal{X}_r^{m\times n}}
\newcommand{\xrr}[2]{\bm{\mathcal{X}}_r^{{#1}\times {#2}}}

\newcommand{\PP}{\bm{\mathcal{P}}_{r}^{\iota}}

\newcommand{\ui}{\mathfrak{I}}
\newcommand{\uimal}{\iota}

\newcommand{\p}{\parallel}

\newcommand{\prv}{\parallel_{rv}}
\newcommand{\si}{\sigma}

\newcommand{\ga}{\gamma}
\newcommand{\Si}{\Sigma}
\newcommand{\La}{\Lambda}
\newcommand{\la}{\lambda}

\newcommand{\de}{\delta}

\newcommand{\non}{\newline\indent}

\newcommand{\no}{\noindent}
\newcommand{\pd}{\no\hspace{2em}{\itshape Proof: }} 
\newcommand{\ds}{\displaystyle}

\newtheorem{theorem}{Theorem}

\newtheorem{remark}{Remark}
\newtheorem{fact}{Fact}

\begin{document}
\title{MV-PURE Spatial Filters with Application to EEG/MEG Source Reconstruction}
\author{Tomasz Piotrowski$^{\star,1,2}$, Jan Nikadon$^{3}$ and David Guti\'{e}rrez$^{4}$,~\IEEEmembership{Senior Member,~IEEE}
  \thanks{The work of TP was supported by a grant from the Polish National Science Centre (UMO-2016/20/W/NZ4/00354).}
  \thanks {The work of JN was supported by a grant from the Polish Ministry of Science and Higher Education  (0094/DIA/2015/44).}\\
  $^1$ Faculty of Physics, Astronomy and Informatics,\\
  Nicolaus Copernicus University,
  Grudziadzka 5/7, 87-100 Torun, Poland\\
  $^2$ Interdisciplinary Center for Modern Technologies,\\
  Nicolaus Copernicus University,
  Wilenska 4, 87-100 Torun, Poland\\
  $^3$ Faculty of Humanities,\\
  Nicolaus Copernicus University,
  Fosa Staromiejska 1A, 87-100 Torun, Poland\\
  $^4$ Center for Research and Advanced Studies,\\ 
  Monterrey's Unit, Apodaca, N.L., 66600, M\'{e}xico\\

\thanks{\textcopyright\ 2019 IEEE.  Personal use of this material is permitted.  Permission from IEEE must be obtained for all other uses, in any current or future media, including reprinting/republishing this material for advertising or promotional purposes, creating new collective works, for resale or redistribution to servers or lists, or reuse of any copyrighted component of this work in other works.}}
  
\maketitle

\begin{abstract}
In this paper we propose spatial filters for a linear regression model which are based on the minimum-variance pseudo-unbiased reduced-rank estimation (MV-PURE) framework. As a sample application, we consider the problem of reconstruction of brain activity from electroencephalographic (EEG) or magnetoencephalographic (MEG) measurements. The proposed filters come in two versions depending on whether or not the EEG/MEG forward model explicitly considers \emph{interfering activity} in the way of brain activity originating in regions different to those of main interest, but measured as correlated with signals of interest by the EEG/MEG sensor array. In both cases, the proposed filters are equipped with a rank-selection criterion minimizing the mean-square error (MSE) of the filter output. Therefore, we consider them as novel nontrivial generalizations of well-known linearly constrained minimum variance (LCMV) and nulling filters. In order to facilitate reproducibility of our research, we provide (jointly with this paper) comprehensive simulation framework that allows for estimation of error of signal reconstruction for a number of spatial filters applied to MEG or EEG signals. Based on this framework, chief properties of proposed filters are verified in a set of detailed simulations.
\end{abstract}

\begin{center} \bfseries EDICS Category: SAM-BEAM, SSP-APPL, SSP-PARE \end{center}
*Corresponding author.
\IEEEpeerreviewmaketitle

\IEEEpeerreviewmaketitle
\section{Introduction} \label{intro}
Beamforming techniques have been used in array signal processing since the seminal paper by Frost \cite{Frost1972}. In electroencephalography (EEG) and magnetoencephalography (MEG), beamforming has been used for signal reconstruction and localization of sources of brain electrical activity. In this field, the dominant approach to solve these problems is to use the linearly constrained minimum variance (LCMV) filter (beamformer), or solutions based on it \cite{VanVeen1997, Gross2001, Sekihara2001, Sekihara2008, Pezeshki2008, Moiseev2011, Diwakar2011}. Indeed, the LCMV filter is implemented in virtually all software enabling EEG/MEG source analysis, e.g., \cite{FieldTrip2011, Brainstorm2011, MNE2014}, and remains useful within EEG/MEG community, e.g., \cite{Keitel2016, Siems2016}. 

However, it has been demonstrated in \cite{VanVeen1997, Sekihara2001, Sekihara2008, Diwakar2011}, and references therein, that the LCMV-based solutions may perform sufficiently well only if certain conditions are satisfied by the EEG/MEG forward model, such as uncorrelatedness of the sources, high signal-to-noise ratio (SNR), sufficiently large spatial separation of sources, or the amount of regularization. Thus, it would be desirable to propose spatial filters that keep the advantages of the LCMV-based filters also in low-SNR regime, especially if EEG/MEG forward model is ill-conditioned. Ideally, such filters would not require heuristic parameter tuning.

Therefore, in this paper we propose a family of reduced-rank filters extending the minimum-variance pseudo-unbiased reduced-rank estimation (MV-PURE) framework \cite{Yamada2006, Piotrowski2008, Piotrowski2009, Yamagishi2013, Piotrowski2014}. Reduced-rank estimators and filters are well-established in signal processing \cite{Brillinger1975, Scharf1991, Stoica1996, Yamashita1996, Scharf1998, Sekihara2001, Sekihara2008, Piotrowski2016b}, as they offer much improved performance compared with full-rank solutions in well-defined settings. Indeed, the proposed filters are solutions of certain mean-square error (MSE) optimization problems and they are equipped with a rank-selection criterion minimizing mean-square error (MSE) of its estimate.

The proposed filters come in two versions depending on whether or not the EEG/MEG forward model explicitly considers \emph{interfering activity} in the way of brain activity originating in regions different to those of main interest, but measured as correlated with signals of interest by the EEG/MEG sensor array. Then, we foresee two possible scenarios: 
\begin{itemize}
\item the interference-free model, in which the proposed filters extend the stochastic MV-PURE approach (previously considered in \cite{Piotrowski2009, Piotrowski2014}) by proposing two new cost functions for stochastic MV-PURE,
\item the model in presence of interference, where the proposed filters extend the nulling filter approach in  \cite{Dalal2006, Hui2010} to the reduced-rank case.
\end{itemize}

In view of the above, the proposed filters are nontrivial generalizations of well-known LCMV and nulling filters parameterized by rank, which is selected according to MSE-minimization principle. As such, the number of applications is essentially the same as any other spatial filter, and is not limited to EEG/MEG settings. In particular, the filters proposed for the model in presence of interference may be specially useful in applications where it is important to have interference removed from the reconstructed activity, e.g., in applications of directed connectivity measures such as partial directed coherence (PDC) \cite{Baccala2001} or directed transfer function (DTF) \cite{Kus2004} as they rely on accurate fitting of the time series representing the sources of interest by means of multivariate autoregressive (MVAR) models.

In order to increase reproducibility of our research, we provide (jointly with this paper) a comprehensive simulation framework that implements signal reconstruction through different spatial filters (including the ones here proposed) to realistic EEG/MEG measurements. Based on this framework, the main properties of proposed filters are verified in a set of detailed simulations. We emphasize that only through simulations one can get a proper assessment of methods' performance, yet our model considers realistic conditions not only for the sources of interest, but also for other interfering brain activity.   

Preliminary short versions of this paper have been presented at conferences \cite{Piotrowski2013, Piotrowski2016a}.

\section{Notation} \label{not}
Assume $\x$ to be a vector of real-valued random variables $x_1,\dots,x_n$, each with a finite variance. The expectation functional is denoted by $\mathbb{E}.$ The norm of $\x$ is defined as\linebreak $\p\x\prv=\sqrt{tr\big(\me{\x\x^t}\big)}$, where $tr$ denotes trace of a certain square matrix. $\bm{I}_n$ stands for identity matrix of size~$n$, while $\bm{0}_{m\times n}$ stands for zero matrix of size $m\times n.$ By $\la_i(\bm{A})$ we denote the $i$-th largest eigenvalue of symmetric matrix $\bm{A}$, by $\si_i(\bm{B})$ the $i$-th largest singular value of matrix $\bm{B}$, by $[\bm{B}]_{u\times v}$ the principal submatrix of $\bm{B}$ composed of the first $u$ rows and $v$ columns of $\bm{B}$, by $rk(\bm{B})$ the rank of $\bm{B}$, and by $[\bm{B}\ \bm{C}]$ the matrix constructed from matrices $\bm{B}$ and $\bm{C}$ by concatenating its columns. Furthermore, we denote by $\bm{B}^\dagger$ the Moore-Penrose pseudoinverse of matrix $\bm{B}$ \cite{Horn1985}. The Frobenius norm of matrix $\bm{B}\in\mn$ is defined as $\p \bm{B}\p_F=\sqrt{tr(\bm{B}\bm{B}^t)}=\sqrt{\sum_{i=1}^m\sum_{j=1}^nb_{i,j}^2}$, where $b_{i,j}$ is the element of $\bm{B}$ at $i$-th row and $j$-th column. We define $\xrr{u}{v}:=\{\bm{X}_r\in\sgen{u}{v}: rk(\bm{X}_r)\leq r\leq\min\{u,v\}\}.$ We denote the orthogonal projection matrix onto subspace $S$ by $\bm{P}_S.$ We assume that vectors considered in this paper are column vectors.

\section{EEG/MEG Measurement Model} \label{assumptions}
The array response (leadfield) matrix defining the relationship between~$l$ dipole sources and $m$ sensors is constructed as $\bm{H}(\th)=\big(H(\theta_1),\dots,H(\theta_l)\big)\in\mathbb{R}^{m \times l}$ where for $i=1,\dots,l$, $H(\theta_i)\in\mathbb{R}^m$ is the leadfield vector of the $i$-th source, $\th=(\theta_1,\dots,\theta_l)$ is such that $\theta_i=\{r_i,u_i\}$, where $r_i$ is the source position, and $u_i$ is the orientation unit vector for the $i$-th source. In this paper we focus on reconstructing activity of sources in predefined locations. Thus, we assume that source positions and orientations are known and fixed during the measurement period. This can be achieved by defining regions of interest using source localization methods, e.g., minimum-norm \cite{Pascual-Marqui1999} or spatial filtering-based methods \cite{Moiseev2011, Piotrowski2014b, Piotrowski2014c}, or referring to neuroscience studies that have identified regions of interest as in \cite{Hui2010}. Nevertheless, the results of this paper apply analogously if the model with unconstrained orientation of the sources is considered. Following \cite{Sekihara2008}, we also assume that the leadfield vectors are linearly independent, which implies in particular that leadfield matrices such as $\bm{H}(\th)$ are of full column rank.

For notational convenience, we will drop the explicit dependence of leadfield matrices on parameter vector $\th$ from now on.

\subsection{Interference-Free Measurement Model} \label{ifm}
Let us consider measurements of brain electrical activity by EEG/MEG sensors at a specified time interval. The random vector $\y\in\sm$ composed of measurements at a given time instant can be modeled as \cite{VanVeen1997, Mosher1999, Sekihara2008},
\begin{equation} \label{model_free}
  \y=\bm{H}\q+\n,
\end{equation}
where $\bm{H}\in\sgen{m}{l}$ is a leadfield matrix of rank $l$ representing leadfields of $l$ sources of interest, $\q\in\sgens{l}$ represents equivalent current (ECD) dipole moments of sources of interest, and $\n\in\sm$ represents remaining brain activity along with noise recorded at the sensors.

Furthermore, we assume that $\q$ and $\n$ are mutually uncorrelated zero-mean weakly stationary stochastic processes. We denote the positive definite covariance matrices of $\q$ and $\n$ as $\bm{Q}$ and $\bm{N}$, respectively. Note that these assumptions imply that $\y$ is also zero-mean weakly stationary processs with positive definite covariance matrix $\bm{R}=\bm{H}\bm{Q}\bm{H}^t+\bm{N}.$ For the noise $\n$, we consider the following model:
\begin{equation} \label{n}
  \n=\bm{H}_b\q_b+\n_m,
\end{equation}
where $\bm{H}_b\q_b\in\mathbb{R}^m$ represents background activity of the brain and~$\n_m\in\mathbb{R}^m$ is the uncorrelated and Gaussian-distributed measurement noise, which includes all the remaining activity of the brain not considered by $\bm{H}_b\q_b.$ The leadfield matrix $\bm{H}_b\in\sgen{m}{p}$ is assumed unknown, and so is the number of sources $p$ representing background activity.

\subsubsection{Spatial Filtering Under MSE Criterion} 
A spatial filter aiming at reconstructing activity $\q$ in (\ref{model_free}) is defined as a matrix $\bm{W}\in\lm$ applied to measurements: 
\begin{equation} \label{rx}
\widehat{\q}=\bm{W}\y.
\end{equation} 
The fidelity of reconstruction is measured using MSE of $\widehat{\q}.$ In terms of (\ref{model_free}), it is expressed as
\begin{multline} \label{mse_free}
\Jn_F(\bm{W})=\\ tr\Big(\me{\big(\bm{W}(\bm{H}\q+\n)-\q\big)\big(\bm{W}(\bm{H}\q+\n)-\q\big)^t}\Big)=\\ tr(\bm{W}\bm{R}\bm{W}^t)-2tr\big(\bm{W}\bm{H}\bm{Q}\big)+c,
\end{multline}
where 
\begin{equation} \label{c}
c=tr(\bm{Q})=\p\q\prv^2>0.
\end{equation}

\subsubsection{LCMV Filter} The most commonly used LCMV spatial filter uses MSE as cost function, and it belongs to a class of filters $\bm{W}^*$ satisfying unit-gain constraint $\bm{W}^*\bm{H}=\bm{I}_l.$ Note that, in view of (\ref{mse_free}), for filters satisfying this constraint we have
\begin{multline} \label{ingen1}
  \Jn_F(\bm{W}^*)=tr(\bm{W}^*\bm{R}(\bm{W}^*)^t)-2tr\big(\bm{W}^*\bm{H}\bm{Q}\big)+c=\\
  tr\big(\bm{W}^*\underbrace{(\bm{H}\bm{Q}\bm{H}^t+\bm{N})}_{\bm{R}}(\bm{W}^*)^t\big)-c=\\ tr(\bm{W}^*\bm{N}(\bm{W}^*)^t).
\end{multline}
Then, the LCMV filter is the solution of the following problem \cite{Frost1972,VanVeen1997}:
\begin{equation} \label{blue}
  \left\{
  \begin{array}{ll}
    \zt{minimize} & tr[\bm{W}\bm{R}\bm{W}^t]\\
    \zt{subject to} & \bm{W}\bm{H}=\bm{I}_l,\\
  \end{array}\right.
\end{equation}
and it has the following unique solution:
\begin{equation} \label{blueish}
  \bm{W}_{LCMV}=(\bm{H}^t\bm{R}^{-1}\bm{H})^{-1}\bm{H}^t\bm{R}^{-1}.
\end{equation}

There are two things that are important to highlight from the previous expressions: first, (\ref{ingen1}) implies that the cost function in (\ref{blue}) can be replaced by $tr(\bm{W}\bm{N}\bm{W}^t)$ in the interference-free case, and thus the LCMV filter may be expressed equivalently in terms of $\bm{N}$ instead of $\bm{R}$ (see \cite{Moiseev2015} for a throughout evaluation of both cases under the presence of modeling and source localization errors); second, since the LCMV filter does not consider the case of correlated interference, then its performance degrades significantly under such condition (see, e.g., \cite{Dalal2006, Hui2010}).

\subsubsection{Conditioning of the Interference-Free Measurement Model} 
From~(\ref{ingen1}), and using the alternative expression of the LCMV filter $\bm{W}_{LCMV}=(\bm{H}^t\bm{N}^{-1}\bm{H})^{-1}\bm{H}^t\bm{N}^{-1}$, we can rewrite the cost function as
\begin{multline} \label{RN_free}
  \Jn_F(\bm{W}_{LCMV})=tr(\bm{W}_{LCMV}\bm{N}\bm{W}_{LCMV}^t)=\\\sum_{i=1}^l\la_i\big((\bm{H}^t\bm{N}^{-1}\bm{H})^{-1}\big).
\end{multline}
Under the simplifying assumption of white $\n$ in (\ref{n}), i.e., assuming that the background activity is spatially uncorrelated at the sensors yielding $\bm{N}=\si^2 \bm{I}_m$, where $\si^2$ is the noise power, we can further rewrite (\ref{RN_free}) as $\la_i\big((\bm{H}^t\bm{N}^{-1}\bm{H})^{-1}\big)=\si^2\la_i\big((\bm{H}^t\bm{H})^{-1}\big)$ and, consequently:
\begin{enumerate}
\item with an increasing level of background activity and/or measurement noise, the MSE of the LCMV filter increases,
\item if $\bm{H}$ has some singular values close to zero, the MSE of the LCMV filter can be in principle arbitrarily large.  
\end{enumerate}
While the first issue is expected, the second has not been fully addressed in the EEG/MEG literature, then in this paper we aim to develop efficient solutions to that problem.

\subsection{Measurement Model in the Presence of Interference} \label{mipofi}
The EEG/MEG measurement model considered in this section expands (\ref{model_free}) by introducing interfering sources exhibiting activity correlated with activity of sources of interest. Namely, we now consider the following \cite{Dalal2006,Hui2010}:
\begin{equation} \label{model}
  \y=\bm{H}_c\q_c+\n,
\end{equation}
where $\bm{H}_c=[\bm{H}\ \bm{H}_I]\in\sgen{m}{(l+k)}$ is a composite leadfield matrix of rank $l+k$ comprised by the originally defined leadfield matrix $\bm{H}\in\ml$ and $\bm{H}_I\in\mk$ which represents leadfields of $k$ interfering sources.\footnote{We note that \cite{Hui2010} uses a slightly different notation of the forward model for interfering sources, where they are considered separately for each region of interest. In our case, they are combined into a single forward model for the interfering sources, yielding forward matrix $\bm{H}_I.$ We took that approach for clarity of the derivation of the proposed filter.} The vector $\q_c=[\q^t\ \q_I^t]^t\in\sgens{l+k}$ is similarly composed of $\q\in\sll$ representing activity of sources of interest, and $\q_I\in\sk$ representing activity of interfering sources.

We also assume that $\q_c$ is a zero-mean weakly stationary stochastic process uncorrelated with $\n$, and its positive definite covariance matrix is given by $\bm{Q}_c.$ These assumptions imply that $\y$ is also zero-mean weakly stationary processs with positive definite covariance matrix $\bm{R}=\bm{H}_c\bm{Q}_c\bm{H}_c^t+\bm{N}.$ 

\subsubsection{Spatial Filtering in the Presence of Interference} 
The MSE of estimate $\widehat{\q}$ of the form (\ref{rx}) is expressed in terms of (\ref{model}) as
\begin{multline} \label{mse}
\Jn_I(\bm{W})=\\ tr\Big(\me{\big(\bm{W}(\bm{H}_c\q_c+\n)-\q\big)\big(\bm{W}(\bm{H}_c\q_c+\n)-\q\big)^t}\Big)=\\ tr(\bm{W}\bm{R}\bm{W}^t)-2tr\big(\bm{W}\bm{H}_c\me{\q_c\q^t}\big)+c,
\end{multline}
where $c$ is given in (\ref{c}).

\subsubsection{Nulling Filter} \label{nl}
The nulling spatial filter proposed in \cite{Dalal2006, Hui2010} extends the LCMV approach by incorporating constraints on the optimization problem which directly remove the impact of correlated interference. It is defined as
\begin{equation} \label{nulling}
  \left\{
  \begin{array}{ll}
    \zt{minimize} & tr[\bm{W}\bm{R}\bm{W}^t]\\
    \zt{subject to} & \left\{
    \begin{array}{l}
      \bm{W}\bm{H}=\bm{I}_l\\
      \bm{W}\bm{H}_I=\bm{0}_{l\times k},\\
    \end{array}\right.\\
  \end{array}\right.
\end{equation}
which has the following unique solution:
\begin{equation} \label{nullingish}
  \bm{W}_{NL}=[\bm{I}_l\ \bm{0}_{l\times k}](\bm{H}_c^t\bm{R}^{-1}\bm{H}_c)^{-1}\bm{H}_c^t\bm{R}^{-1}.
\end{equation}
In view of (\ref{mse}), the MSE of filters $\bm{W}^{**}$ that satisfy the constraints of the optimization problem in (\ref{nulling}) is given by
\begin{multline} \label{ingen2}
  \Jn_I(\bm{W}^{**})= tr(\bm{W}^{**}\bm{R}(\bm{W}^{**})^t)-2tr\big(\bm{W}^{**}\bm{H}_c\me{\q_c\q^t}\big)+c=\\
  tr\big(\bm{W}^{**}\underbrace{(\bm{H}_c\bm{Q}_c\bm{H}_c^t+\bm{N})}_{\bm{R}}(\bm{W}^{**})^t\big)-c=\\ tr(\bm{W}^{**}\bm{N}(\bm{W}^{**})^t).
\end{multline}

Again, note that (\ref{ingen2}) implies that the cost function in (\ref{nulling}) can be replaced by $tr(\bm{W}\bm{N}\bm{W}^t)$ for the nulling filter, and thus the nulling filter may be expressed equivalently in terms of $\bm{N}$ instead of $\bm{R}$, just as the LCMV filter in the interference-free case. The nulling filter is relevant as it has showed remarkable improvement in reconstruction performance over the LCMV filter in the presence of interfering sources \cite{Dalal2006,Hui2010}.

\subsubsection{Conditioning of the Measurement Model in the Presence of Interference}
Using the alternative expression of the nulling filter $\bm{W}_{NL}=[\bm{I}_l\ \bm{0}_{l\times k}](\bm{H}_c^t\bm{N}^{-1}\bm{H}_c)^{-1}\bm{H}_c^t\bm{N}^{-1}$, from (\ref{ingen2}) and similarly as for the LCMV filter, we have that
\begin{multline} \label{RN2}  \Jn_I(\bm{W}_{NL})=tr(\bm{W}_{NL}\bm{N}\bm{W}_{NL}^t)=\\\sum_{i=1}^l\la_i\Big(\big[(\bm{H}_c^t\bm{N}^{-1}\bm{H}_c)^{-1}\big]_{l\times l}\Big).
\end{multline}
While this expression cannot be analyzed as easily as (\ref{RN_free}), the key observation is that, in the presence of interference, the value of (\ref{RN2}) for the nulling filter will be at least as large as the corresponding value of (\ref{RN_free}) for the LCMV filter in the interference-free case. This is the trade-off of further constraining the optimization problem in (\ref{nulling}) compared with (\ref{blue}). In particular, the sensitivity of the nulling filter to the ill-conditioning of the forward model is essentially the same as of the LCMV filter.

\subsection{Challenging the Ill-Conditioning of the Measurement Model} \label{Ch1}
One efficient way to lower sensitivity on ill-conditioning is to reduce the rank of the filter \cite{Scharf1991, Hua2001, Gutierrez2006,  Piotrowski2016b}, which is the approach we take in this paper. Specifically, we introduce a parameter $r$ such that $1\leq r\leq l$ is the rank of a reduced-rank filter $\bm{W}_r\in\xrr{l}{m}.$ The reduced-rank approach relaxes unit-gain constraint $\bm{W}\bm{H}=\bm{I}_l$ of the LCMV and nulling filters, as the exact equality cannot be obtained in this case. Thus, in order to employ the corresponding constraint in the reduced-rank case, we take advantage of the MV-PURE framework \cite{Yamada2006, Piotrowski2008, Piotrowski2009} and insist that the reduced-rank filter is designed to introduce the smallest deviation from unit-gain condition among filters of rank at most $r.$ More precisely, the proposed reduced-rank filter should satisfy:
\begin{equation} \label{argmin}
  \bm{W}_r^\bullet\in\PP:=\arg\min_{\bm{W}_r\in\xrr{l}{m}}\p \bm{W}_r\bm{H}-\bm{I}_l\p_\uimal^2,\ \uimal\in\ui,
\end{equation}
for all $\uimal$, where $\ui$ is the index set of all unitarily invariant norms.

In essence, the above approach introduces parameter $r$ - the rank of the matrix -, hoping that it can be selected to optimize certain cost function. As it will be shown in the next section, it is possible to select $r$ such that the MSE of the resulting filter is minimized. In fact, it has been previously shown that this relaxation of unit-gain constraint deals efficiently with ill-conditioning, and it may even achieve lower MSE than their full-rank counterparts \cite{Piotrowski2014,Piotrowski2016b}.

\section{Proposed Spatial Filters in Presence of Interference} \label{proposed}
In this and the following section, we introduce the proposed filters which use the reduced-rank MV-PURE paradigm embodied in constraint (\ref{argmin}). Unlike in Sections \ref{ifm} and \ref{mipofi}, we begin with the case where the interfering sources are explicitly considered since the proofs of theorems establishing filters for interference-free model can be obtained as special cases of the ones in presence of interference.

\subsection{Optimization Problems}
We propose a new filter as a solution of the following optimization problem parameterized by $r$:
\begin{equation} \label{rr_nulling}
  \left\{
  \begin{array}{ll}
    \zt{minimize} & \Jn_I(\bm{W}_r)\\
    \zt{subject to} & \left\{
    \begin{array}{l}
      \bm{W}_r\in\bigcap_{\uimal\in\ui}\PP\\
      \bm{W}_r\bm{H}_I=\bm{0}_{l\times k},\\
    \end{array}\right.\\
  \end{array}\right.
\end{equation}
where $\PP$ is defined in (\ref{argmin}).

Note that the relaxation of unit-gain constraint $\bm{W}\bm{H}=\bm{I}_l$ used in (\ref{nulling}) implies that equalities analogous to those in (\ref{ingen2}) do not hold for (\ref{rr_nulling}). For this reason, we introduce two additional versions of the proposed filter with cost functions $tr(\bm{W}_r\bm{R}\bm{W}_r^t)$ and $tr(\bm{W}_r\bm{N}\bm{W}_r^t)$, respectively, and the same constraints on $\bm{W}_r$ as in (\ref{rr_nulling}), i.e.,
\begin{equation} \label{rr_nulling_R}
  \left\{
  \begin{array}{ll}
    \zt{minimize} & tr(\bm{W}_r\bm{R}\bm{W}_r^t)\\
    \zt{subject to} & \left\{
    \begin{array}{l}
      \bm{W}_r\in\bigcap_{\uimal\in\ui}\PP\\
      \bm{W}_r\bm{H}_I=\bm{0}_{l\times k},\\
    \end{array}\right.\\
  \end{array}\right.
\end{equation}
and
\begin{equation} \label{rr_nulling_N}
  \left\{
  \begin{array}{ll}
    \zt{minimize} & tr(\bm{W}_r\bm{N}\bm{W}_r^t)\\
    \zt{subject to} & \left\{
    \begin{array}{l}
      \bm{W}_r\in\bigcap_{\uimal\in\ui}\PP\\
      \bm{W}_r\bm{H}_I=\bm{0}_{l\times k},\\
    \end{array}\right.\\
  \end{array}\right.
\end{equation}
where $\PP$ is defined in (\ref{argmin}). Together, these three cost functions produce three different filters, unlike in the case of nulling filter, where all these cost functions are equivalent.

Based on the proposed filters (\ref{rr_nulling})-(\ref{rr_nulling_N}), our approach is then to select rank $r$ which minimizes MSE. As it will be demonstrated in the next section, this strategy can be efficiently applied to all three versions of the filter produced by the three cost functions considered. In this way, the proposed filters are nontrivial generalizations of the nulling filter. 

\subsection{Closed Algebraic Forms}
Theorem \ref{Th1} below establishes closed algebraic form of the proposed filter obtained as the solution of the optimization problem (\ref{rr_nulling}). Then, Theorems \ref{Th2} and \ref{Th3} present closed algebraic forms of the proposed filter for the alternative cost functions $tr(\bm{W}_r\bm{R}\bm{W}_r^t)$ and $tr(\bm{W}_r\bm{N}\bm{W}_r^t)$, respectively.

\begin{theorem} \label{Th1}
  For a given $r$ such that $1\leq r\leq l$, the solution to optimization problem (\ref{rr_nulling}) is given by
  \begin{equation} \label{sl}
    \bm{W}_r^\star=\bm{P}_{\bm{K}_r^{(1)}}\big(\bm{P}_{\rangeperp{\bm{G}_I}}\bm{G}\big)^{\dagger}\bm{P}_{\rangeperp{\bm{G}_I}}\bm{R}^{-1/2},
  \end{equation}
  where
  \begin{itemize}
  \item $\bm{R}^{-1/2}$ is the unique positive definite matrix satisfying $\bm{R}^{-1/2}\bm{R}^{-1/2}=\bm{R}^{-1}$ \cite{Horn1985},
  \item $\bm{G}:=\bm{R}^{-1/2}\bm{H}$ and $\bm{G}_I:=\bm{R}^{-1/2}\bm{H}_I$,
  \item $\bm{P}_{\rangeperp{\bm{G}_I}}$ is the orthogonal projection matrix onto orthogonal complement of range of $\bm{G}_I$,
  \item $\bm{P}_{\bm{K}_r^{(1)}}$ is the orthogonal projection matrix onto subspace spanned by eigenvectors corresponding to $\de_1^{(1)}\leq\dots\leq\de_r^{(1)}$ - the $r$ smallest eigenvalues of the symmetric matrix
\begin{equation} \label{K1}
\bm{K}^{(1)}=(\bm{P}_{\rangeperp{\bm{G}_I}}\bm{G})^\dagger \bm{P}_{\rangeperp{\bm{G}_I}}\big((\bm{P}_{\rangeperp{\bm{G}_I}}\bm{G})^\dagger\big)^t-2\bm{Q}.
\end{equation}  
  \end{itemize}
  Its corresponding MSE is given by
\begin{equation} \label{J1}
\Jn_I(\bm{W}_r^\star)=tr(\bm{P}_{\bm{K}_r^{(1)}}\bm{K}^{(1)})+c=\sum_{i=1}^r\de_i^{(1)}+c,
\end{equation}
where $c=tr(\bm{Q}).$
\end{theorem}
\pd See Appendix \ref{pdTh1}.

\begin{theorem} \label{Th2}
With notation as in Theorem \ref{Th1}, for a given $r$ such that $1\leq r\leq l$, the solution to optimization problem (\ref{rr_nulling_R}) is given by
\begin{equation} \label{sl2}
\bm{W}_r^\star=\bm{P}_{\bm{K}_r^{(2)}}\big(\bm{P}_{\rangeperp{\bm{G}_I}}\bm{G}\big)^{\dagger}\bm{P}_{\rangeperp{\bm{G}_I}}\bm{R}^{-1/2},
\end{equation}
where $\bm{P}_{\bm{K}_r^{(2)}}$ is the orthogonal projection matrix onto subspace spanned by eigenvectors corresponding to $\de_1^{(2)}\leq\dots\leq\de_r^{(2)}$ - the $r$ smallest eigenvalues of the symmetric matrix
\begin{equation}
  \bm{K}^{(2)}=\bm{K}^{(1)}+2\bm{Q}.
\end{equation}  
The MSE for this solution can be written as
\begin{equation} \label{J2}
\Jn_I(\bm{W}_r^\star)=tr(\bm{P}_{\bm{K}_r^{(2)}}\bm{K}^{(1)})+c=\sum_{i=1}^r\de_i^{(2)}-2tr(\bm{P}_{\bm{K}_r^{(2)}}\bm{Q})+c,
\end{equation}
where $c=tr(\bm{Q}).$
\end{theorem}
\pd See Appendix \ref{pdTh2}.

\begin{theorem} \label{Th3}
For a given $r$ such that $1\leq r\leq l$, the solution to optimization problem (\ref{rr_nulling_N}) is given by:
\begin{equation} \label{sl3}
\bm{W}_r^\star=\bm{P}_{\bm{K}_r^{(3)}}\big(\bm{P}_{\rangeperp{\bm{F}_I}}\bm{F}\big)^{\dagger}\bm{P}_{\rangeperp{\bm{F}_I}}\bm{N}^{-1/2},
\end{equation}
  where
  \begin{itemize}
  \item $\bm{N}^{-1/2}$ is the unique positive definite matrix satisfying $\bm{N}^{-1/2}\bm{N}^{-1/2}=\bm{N}^{-1}$,
  \item $\bm{F}:=\bm{N}^{-1/2}\bm{H}$ and $\bm{F}_I:=\bm{N}^{-1/2}\bm{H}_I$,
  \item $\bm{P}_{\rangeperp{\bm{F}_I}}$ is the orthogonal projection matrix onto orthogonal complement of range of $\bm{F}_I$,
  \item $\bm{P}_{\bm{K}_r^{(3)}}$ is the orthogonal projection matrix onto subspace spanned by eigenvectors corresponding to $\de_1^{(3)}\leq\dots\leq\de_r^{(3)}$ - the $r$ smallest eigenvalues of the symmetric matrix
\begin{equation}
\bm{K}^{(3)}=(\bm{P}_{\rangeperp{\bm{F}_I}}\bm{F})^\dagger \bm{P}_{\rangeperp{\bm{F}_I}}\big((\bm{P}_{\rangeperp{\bm{F}_I}}\bm{F})^\dagger\big)^t.
\end{equation}  
  \end{itemize}
The MSE of (\ref{sl3}) is given by
\begin{equation} \label{J3}
\Jn_I(\bm{W}_r^\star)=tr(\bm{P}_{\bm{K}_r^{(3)}}\bm{K}^{(4)})+c=\sum_{i=1}^r\de_i^{(3)}-tr(\bm{P}_{\bm{K}_r^{(3)}}\bm{Q})+c,
\end{equation}
where $c=tr(\bm{Q})$ and
\begin{equation}
  \bm{K}^{(4)}=\bm{K}^{(3)}-\bm{Q}.
\end{equation}
\end{theorem}
\pd See Appendix \ref{pdTh3}.

There are some observations with regards to the results of Theorems \ref{Th1}-\ref{Th3} that deserve to be highlighted:
\begin{remark} \label{VIPR1}
\begin{itemize}
\item For no rank constraint, i.e., $r=l$, and based on Theorems \ref{Th1}-\ref{Th3}, the nulling filter (\ref{nullingish}) is alternatively written as
\begin{multline} \label{WNL_new}
  \bm{W}_{NL}=\big(\bm{P}_{\rangeperp{\bm{G}_I}}\bm{G}\big)^{\dagger}\bm{P}_{\rangeperp{\bm{G}_I}}\bm{R}^{-1/2}=\\
  \big(\bm{P}_{\rangeperp{\bm{F}_I}}\bm{F}\big)^{\dagger}\bm{P}_{\rangeperp{\bm{F}_I}}\bm{N}^{-1/2}.
\end{multline}
\item In view of (\ref{WNL_new}), the filters in (\ref{sl}), (\ref{sl2}) and (\ref{sl3}) differ only in the subspace that the estimate of the nulling filter is orthogonally projected onto. Namely, they are of the form:
\begin{equation} \label{elegant1}
\bm{W}_r^\star=\bm{P}_{\bm{K}_r^{(i)}}\bm{W}_{NL},\ i=1,2,3.
\end{equation}
\item Furthermore, from (\ref{WNL_new}), and using the fact that orthogonal projection matrices are both symmetric and idempotent \cite{Horn1985}, we may obtain alternative representations of matrices $\bm{K}^{(1)}$ and $\bm{K}^{(3)}$ as:
  \begin{equation}
\bm{K}^{(1)}=\bm{W}_{NL}\bm{R}\bm{W}_{NL}^t-2\bm{Q},
  \end{equation}
  \begin{equation}
\bm{K}^{(3)}=\bm{W}_{NL}\bm{N}\bm{W}_{NL}^t,
  \end{equation}
  with $\bm{K}^{(2)}=\bm{K}^{(1)}+2\bm{Q}=\bm{W}_{NL}\bm{R}\bm{W}_{NL}^t$ and $\bm{K}^{(4)}=\bm{K}^{(3)}-\bm{Q}=\bm{W}_{NL}\bm{N}\bm{W}_{NL}^t-\bm{Q}.$
\item It is seen from the previous two points that the proposed filters, if cast in the form (\ref{elegant1}), require computation of the nulling filter along with the orthogonal projection onto a minor subspace of one of $\bm{K}^{(i)}\in\mathbb{R}^{l\times l}$ matrices for $i=1,2,3$, whose complexity is $\mathcal{O}(l^3)$ using na\"ive implementation~\cite{Golub1996}. Thus, the computational effort needed to compute the proposed filters in the form (\ref{elegant1}) is of the same order as of the nulling filter. We should also note that typically $l\ll m$ in EEG/MEG source reconstruction.
\item The filter (\ref{sl}) is expressed in terms of $\bm{Q}$ through $\bm{K}^{(1)}$, whereas the filters in (\ref{sl2}) and (\ref{sl3}) do not depend explicitly on $\bm{Q}.$
\item $\bm{Q}$ may be estimated as $\widehat{\bm{Q}}=[\widehat{\bm{Q}}_c]_{l\times l}$, where $\widehat{\bm{Q}}_c$ is obtained through \cite[Lemma 1]{Piotrowski2014}
\begin{equation} \label{L1_2014}
  \bm{Q}_c=(\bm{H}_c^t\bm{R}^{-1}\bm{H}_c)^{-1}-(\bm{H}_c^t\bm{N}^{-1}\bm{H}_c)^{-1},
\end{equation}
in which $\bm{R}$ and $\bm{N}$ are replaced by the finite sample estimates.
\item The MSE in (\ref{J1}), (\ref{J2}) and (\ref{J3}) yield a natural rank-selection method applicable to all forms of the proposed filter by selecting $r$ which minimizes them. We note \emph{en passant} that (\ref{J1}) depends on $r$ only through eigenvalues of the symmetric matrix $\bm{K}^{(1)}$, while (\ref{J2}) depends on $r$ through eigenvalues of the positive semidefinite matrix $\bm{K}^{(2)}$ and $tr(\bm{P}_{\bm{K}_r^{(2)}}\bm{Q}).$ Similarly, (\ref{J3}) depends on $r$ through eigenvalues of the positive semidefinite matrix $\bm{K}^{(3)}$ and $tr(\bm{P}_{\bm{K}_r^{(3)}}\bm{Q}).$
\item Finally, as seen in \cite{Moiseev2015} for the LCMV filter, both the nulling and the proposed filters are expected to show differences in performance whether $\bm{R}$ or $\bm{N}$ is used in the presence of modeling and source localization errors. We also note that \cite{Hui2010} considers only the expression of the nulling filter in terms of $\bm{R}$, but not $\bm{N}.$ 
\end{itemize}
\end{remark}

\subsection{Interactions Among Sources Based on MVAR Model} \label{interactions}
One of possible applications of the nulling filter and the filters proposed in this section is in identifying interactions among sources of interest by means of measuring causal dependencies among their reconstructed activities, as discussed in \cite{Hui2010,Haufe2013}. In particular, in \cite{Hui2010} the authors have used the partial directed coherence (PDC) \cite{Baccala2001} of directed causal dependencies based on fitting multivariate autoregressive (MVAR) model to the reconstructed activity of sources of interest.

As the MVAR model is inherently sensitive to linear mixing among time series considered, it is important to obtain the estimate of activity of sources of interest which is to the largest possible extent free from contamination of interfering sources. Indeed, both nulling and filters proposed in this section produce interference-free estimates, as from (\ref{nullingish}), (\ref{sl}), (\ref{sl2}) and (\ref{sl3}) we have that:
\begin{multline} \label{atii1}
  \bm{W}_{NL}\y=\bm{W}_{NL}(\bm{H}_c\q_c+\n)=\\ \bm{W}_{NL}\big([\bm{H}\ \bm{H}_I][\q^t\ \q_I^t]^t+\bm{H}_b\q_b+\n_m\big)=\\
  \q+\underbrace{\bm{W}_{NL}\bm{H}_b\q_b+\bm{W}_{NL}\n_m}_{\zt{reconstructed noise}},
\end{multline}  
and
\begin{multline} \label{atii2}
\bm{W}^\star_r\y=\bm{W}^\star_r(\bm{H}_c\q_c+\n)=\\ \bm{W}^\star_r\big([\bm{H}\ \bm{H}_I][\q^t\ \q_I^t]^t+\bm{H}_b\q_b+\n_m\big)=\\ \bm{P}_{\bm{K}_r^{(i)}}\q+\underbrace{\bm{W}^\star_r\bm{H}_b\q_b+\bm{W}^\star_r\n_m}_{\zt{reconstructed noise}},\ i=1,2,3,
\end{multline}  
respectively. Then, (\ref{atii1}) and (\ref{atii2}) show that the value of $r$ selected to minimize the MSE of $\bm{W}^\star_r\y$ (which includes $\bm{W}_{NL}\y$ as a special case for $r=l$) introduces a trade-off between the dimension of subspace the signal of interest $\q$ is orthogonally projected onto and the power of the reconstructed noise. Thus, if $\q$ can  be well-fit into $r$-dimensional subspace, we shall expect more accurate MVAR model fitted to the reconstructed activity due to efficient suppression of noise reconstructed by the filter of rank $r$ compared to the full-rank nulling filter. 

Clearly, the above analysis could be used for other applications of the nulling and proposed filters as well. However, their effectiveness depends on the assumption that all interfering sources have been identified and correctly localized. This may not always be the case in practice, as the following section will describe.

\subsection{Extension to Patch Constraints}
As discussed in \cite{Hui2010}, in certain scenarios it may be difficult to determine exact locations of interfering brain sources. Instead, whole patches of cortical and subcortical regions may be considered as contributing correlated activity to the measured signal $\y.$ Then, a complete removal of its impact on reconstructed activity would require imposing constraints $\bm{W}\bm{H}_I=\bm{0}_{l\times k}$ for a large number of interfering sources $k.$ This would result in very tight constraints on the optimization problem in (\ref{nulling}) and, consequently, large MSE of the resulting filter.

For such cases, a heuristic solution proposed in \cite{Hui2010} is to increase the number of degrees of freedom available for the filter by relaxing constraint $\bm{W}\bm{H}_I=\bm{0}_{l\times k}.$ This is achieved by replacing the leadfield matrix of interfering sources $\bm{H}_I$ with its best approximation $\bm{H}_{I_s}$ of rank $s\leq k.$ More precisely, the nulling constraints are replaced with so-called patch constraints \cite{Hui2010}
\begin{equation} \label{nulling_relaxed}
\bm{W}\bm{H}_{I_s}=\bm{0}_{l\times k},
\end{equation}
where $s$ is a rank of the approximation such that $1\leq s\leq k.$ This relaxation reduces the number of degrees of freedom needed for nulling constraints from $k$ to $s$ at the price of having them only approximately enforced.

In fact, using patch constraints (\ref{nulling_relaxed}) in place of nulling constraints $\bm{W}\bm{H}_I=\bm{0}_{l\times k}$ yields significant differences to the behaviour of the nulling filter introduced in Section \ref{nl} and filters proposed in this section. Specifically, if we consider patch constraints (\ref{nulling_relaxed}) in place of nulling constraints $\bm{W}\bm{H}_I=\bm{0}_{l\times k}$ for the nulling filter and the filters introduced in Theorems \ref{Th1}-\ref{Th3}, then we can pose the following optimization problems:
    \begin{equation} \label{nulling_R_ok}
  \left\{
  \begin{array}{ll}
    \zt{minimize} & tr[\bm{W}\bm{R}\bm{W}^t]\\
    \zt{subject to} & \left\{
    \begin{array}{l}
      \bm{W}\bm{H}=\bm{I}_l\\
      \bm{W}\bm{H}_{I_s}=\bm{0}_{l\times k},\\
    \end{array}\right.\\
  \end{array}\right. 
\end{equation}
    and
    \begin{equation} \label{nulling_N_ok}
  \left\{
  \begin{array}{ll}
    \zt{minimize} & tr[\bm{W}\bm{N}\bm{W}^t]\\
    \zt{subject to} & \left\{
    \begin{array}{l}
      \bm{W}\bm{H}=\bm{I}_l\\
      \bm{W}\bm{H}_{I_s}=\bm{0}_{l\times k},\\
    \end{array}\right.\\
  \end{array}\right.
\end{equation}
which now yield two different filters. This is due to $\bm{W}^{**_s}$ satisfying constraints of these optimization problems, hence we have that $\bm{W}^{**_s}\bm{H}_c=\bm{W}^{**_s}[\bm{H}\ \bm{H}_I]\neq [\bm{I}_l\ \bm{0}_{l\times k}]$, thus in particular, the third equality in (\ref{ingen2}) does not hold.

Furthermore, the proof of Theorem \ref{Th1} no longer holds at (\ref{mvptrick}), because it turns out that
\begin{equation} \label{mvptrick_evil}
\bm{Z}_r\bm{G}_c\me{\q_c\q^t}=\bm{Z}_r[\bm{G}\ \bm{G}_I]\me{\q_c\q^t}\neq [\bm{Z}_r\bm{G}\ \bm{0}_{l\times k}]\me{\q_c\q^t}. 
\end{equation}
Therefore, the optimization problem (\emph{cf}. (\ref{rr_nulling}))
\begin{equation} \label{rr_nulling_evil}
  \left\{
  \begin{array}{ll}
    \zt{minimize} & \Jn_I(\bm{W}_r)\\
    \zt{subject to} & \left\{
    \begin{array}{l}
      \bm{W}_r\in\bigcap_{\uimal\in\ui}\PP\\
      \bm{W}_r\bm{H}_{I_s}=\bm{0}_{l\times k},\\
    \end{array}\right.\\
  \end{array}\right.
\end{equation}
with $\PP$ defined in (\ref{argmin}), cannot be solved using methods provided in the proof of Theorem \ref{Th1}. For the same reason, the MSE cost function $J_I(\bm{W}_r)$ cannot be evaluated exactly for the solutions of the following optimization problems (\emph{cf}. (\ref{rr_nulling_R}) and (\ref{rr_nulling_N})):
\begin{equation} \label{rr_nulling_R_ok}
  \left\{
  \begin{array}{ll}
    \zt{minimize} & tr(\bm{W}_r\bm{R}\bm{W}_r^t)\\
    \zt{subject to} & \left\{
    \begin{array}{l}
      \bm{W}_r\in\bigcap_{\uimal\in\ui}\PP\\
      \bm{W}_r\bm{H}_{I_s}=\bm{0}_{l\times k},\\
    \end{array}\right.\\
  \end{array}\right.
\end{equation}
and
\begin{equation} \label{rr_nulling_N_ok}
  \left\{
  \begin{array}{ll}
    \zt{minimize} & tr(\bm{W}_r\bm{N}\bm{W}_r^t)\\
    \zt{subject to} & \left\{
    \begin{array}{l}
      \bm{W}_r\in\bigcap_{\uimal\in\ui}\PP\\
      \bm{W}_r\bm{H}_{I_s}=\bm{0}_{l\times k}.\\
    \end{array}\right.\\
  \end{array}\right.
\end{equation}

Nevertheless, closed algebraic forms of solutions of optimization problems (\ref{rr_nulling_R_ok}) and (\ref{rr_nulling_N_ok}) can be obtained, as the following theorem shows.
  \begin{theorem} \label{ThP}
    With notation as in Theorems \ref{Th2} and~\ref{Th3}, the closed algebraic forms of optimization problems (\ref{rr_nulling_R_ok}) and (\ref{rr_nulling_N_ok}) are, respectively
    \begin{equation} \label{sl2P}
\bm{W}_r^\star=\bm{P}_{\bm{K}_r^{(2)}}\big(\bm{P}_{\rangeperp{\bm{G}_{I_s}}}\bm{G}\big)^{\dagger}\bm{P}_{\rangeperp{\bm{G}_{I_s}}}\bm{R}^{-1/2},
    \end{equation}
    and
    \begin{equation} \label{sl3P}
\bm{W}_r^\star=\bm{P}_{\bm{K}_r^{(3)}}\big(\bm{P}_{\rangeperp{\bm{F}_{I_s}}}\bm{F}\big)^{\dagger}\bm{P}_{\rangeperp{\bm{F}_{I_s}}}\bm{N}^{-1/2},
\end{equation}
where $\bm{G}_{I_s}:=\bm{R}^{-1/2}\bm{H}_{I_s}$ and $\bm{F}_{I_s}:=\bm{N}^{-1/2}\bm{H}_{I_s}.$ In particular, for $r=l$, the solutions to optimization problems 
(\ref{nulling_R_ok}) and (\ref{nulling_N_ok}) are obtained as
    \begin{equation} \label{sl2Pnl}
\bm{W}_r^\star=\big(\bm{P}_{\rangeperp{\bm{G}_{I_s}}}\bm{G}\big)^{\dagger}\bm{P}_{\rangeperp{\bm{G}_{I_s}}}\bm{R}^{-1/2},
    \end{equation}
    and
    \begin{equation} \label{sl3Pnl}
\bm{W}_r^\star=\big(\bm{P}_{\rangeperp{\bm{F}_{I_s}}}\bm{F}\big)^{\dagger}\bm{P}_{\rangeperp{\bm{F}_{I_s}}}\bm{N}^{-1/2},
\end{equation}
respectively.    
  \end{theorem}
  \pd From Fact \ref{sat} given in the Appendix \ref{kru} one has that $\bm{H}_{I_s}=[\bm{U}_{\bm{H}_I}]_{m\times s}[\bm{\Si}_{\bm{H}_I}]_{s\times s}([\bm{V}_{\bm{H}_I}]_{k\times s})^t$, where $\bm{U}_{\bm{H}_I}\bm{\Si}_{\bm{H}_I}\bm{V}_{\bm{H}_I}^t$ is a singular value decomposition of $\bm{H}_I.$ Therefore, $\range{\bm{H}_{I_s}}\subset\range{\bm{H}_I}$, and consequently also $\range{\bm{G}_{I_s}}\subset\range{\bm{G}_I}$ and $\range{\bm{F}_{I_s}}\subset\range{\bm{F}_I}.$ Hence, $\range{\bm{G}}\cap\range{\bm{G}_{I_s}}=\{0\}$ and $\range{\bm{F}}\cap\range{\bm{F}_{I_s}}=\{0\}.$ On this basis, the parts of the proofs of Theorems \ref{Th2} and \ref{Th3} needed to obtain closed algebraic form can be reproduced to obtain (\ref{sl2P}) and (\ref{sl3P}), respectively. Expressions (\ref{sl2Pnl}) and (\ref{sl3Pnl}) follow from (\ref{sl2P}) and (\ref{sl3P}), respectively, as special cases for $r=l.$ \IEEEQEDhere

In the case of patch-constraints, we propose to use the filters as given in Theorem \ref{ThP}, then the rank of the filter is selected to approximately minimize its output MSE. This is achieved by using expressions (\ref{J2}) of Theorem \ref{Th2} and (\ref{J3}) of Theorem~\ref{Th3}, replacing $\bm{G}_I$ with $\bm{G}_{I_s}$ in the former case and $\bm{F}_I$ with $\bm{F}_{I_s}$ in the latter. We will employ this approach in the numerical simulations considered in Section \ref{ne}.

\section{Proposed Spatial Filters For Interference-Free Model} \label{proposed_free}
Correspondingly to optimization problems (\ref{rr_nulling})-(\ref{rr_nulling_N}) considered in Section \ref{proposed}, we propose the following filters as solutions of the optimization problems parameterized by $r$ for the case of the interference-free model in (\ref{model_free}):
\begin{equation} \label{rr_nulling_free}
  \left\{
  \begin{array}{ll}
    \zt{minimize} & \Jn_F(\bm{W}_r)\\
    \zt{subject to} & \bm{W}_r\in\bigcap_{\uimal\in\ui}\PP,\\    
  \end{array}\right.
\end{equation}
\begin{equation} \label{rr_nulling_free_R}
  \left\{
  \begin{array}{ll}
    \zt{minimize} & tr(\bm{W}_r\bm{R}\bm{W}_r^t)\\
    \zt{subject to} & \bm{W}_r\in\bigcap_{\uimal\in\ui}\PP,\\    
  \end{array}\right.
\end{equation}
\begin{equation} \label{rr_nulling_free_N}
  \left\{
  \begin{array}{ll}
    \zt{minimize} & tr(\bm{W}_r\bm{N}\bm{W}_r^t)\\
    \zt{subject to} & \bm{W}_r\in\bigcap_{\uimal\in\ui}\PP,\\    
  \end{array}\right.
\end{equation}
where $\PP$ is defined in (\ref{argmin}).

The following theorem establishes the closed algebraic forms of filters defined through (\ref{rr_nulling_free})-(\ref{rr_nulling_free_N}).
\begin{theorem} \label{Th4}
  With notation as in Theorems \ref{Th1}-\ref{Th3}:
  \begin{enumerate}
    \item For a given $r$ such that $1\leq r\leq l$, the solution to optimization problem (\ref{rr_nulling_free}) is given by
  \begin{equation} \label{sl_free}
    \bm{W}_r^\star=\bm{P}_{\bm{L}_r^{(1)}}\bm{G}^{\dagger}\bm{R}^{-1/2},
  \end{equation}
  where $\bm{P}_{\bm{L}_r^{(1)}}$ is the orthogonal projection matrix onto subspace spanned by eigenvectors corresponding to $\ga_1^{(1)}\leq\dots\leq\ga_r^{(1)}$ - the $r$ smallest eigenvalues of the symmetric matrix
\begin{equation} \label{L1}
\bm{L}^{(1)}=\bm{G}^\dagger(\bm{G}^\dagger)^t-2\bm{Q}.
\end{equation}  
  Then, the MSE is given by
\begin{equation} \label{J1_free}
\Jn_F(\bm{W}_r^\star)=tr(\bm{P}_{\bm{L}_r^{(1)}}\bm{L}^{(1)})+c=\sum_{i=1}^r\ga_i^{(1)}+c,
\end{equation}
where $c=tr(\bm{Q}).$

\item The solution to optimization problem (\ref{rr_nulling_free_R}) is given by
  \begin{equation} \label{sl2_free}
    \bm{W}_r^\star=\bm{P}_{\bm{L}_r^{(2)}}\bm{G}^{\dagger}\bm{R}^{-1/2},
  \end{equation}
  where $\bm{P}_{\bm{L}_r^{(2)}}$ is the orthogonal projection matrix onto subspace spanned by eigenvectors corresponding to $\ga_1^{(2)}\leq\dots\leq\ga_r^{(2)}$ - the $r$ smallest eigenvalues of the symmetric matrix
\begin{equation} \label{L2}
\bm{L}^{(2)}=\bm{L}^{(1)}+2\bm{Q}=\bm{G}^\dagger(\bm{G}^\dagger)^t.
\end{equation}  
  In this case, the MSE is given by
\begin{multline} \label{J2_free}
\Jn_F(\bm{W}_r^\star)=tr(\bm{P}_{\bm{L}_r^{(2)}}\bm{L}^{(1)})+c=\\\sum_{i=1}^r\ga_i^{(2)}-2tr(\bm{P}_{\bm{L}_r^{(2)}}\bm{Q})+c.
\end{multline}

\item The solution to optimization problem (\ref{rr_nulling_free_N}) is given by
  \begin{equation} \label{sl3_free}
    \bm{W}_r^\star=\bm{P}_{\bm{L}_r^{(3)}}\bm{F}^{\dagger}\bm{N}^{-1/2},
  \end{equation}
  where $\bm{P}_{\bm{L}_r^{(3)}}$ is the orthogonal projection matrix onto subspace spanned by eigenvectors corresponding to $\ga_1^{(3)}\leq\dots\leq\ga_r^{(3)}$ - the $r$ smallest eigenvalues of the symmetric matrix
\begin{equation} \label{L3}
\bm{L}^{(3)}=\bm{F}^\dagger(\bm{F}^\dagger)^t.
\end{equation}  
  In this case, the MSE is given by
\begin{multline} \label{J3_free}
\Jn_F(\bm{W}_r^\star)=tr(\bm{P}_{\bm{L}_r^{(3)}}\bm{L}^{(4)})+c=\\\sum_{i=1}^r\ga_i^{(3)}-tr(\bm{P}_{\bm{L}_r^{(3)}}\bm{Q})+c,
\end{multline}
where
\begin{equation}
\bm{L}^{(4)}=\bm{L}^{(3)}-\bm{Q}=\bm{F}^\dagger(\bm{F}^\dagger)^t-\bm{Q}.
\end{equation}
\end{enumerate}
\end{theorem}  
\pd The proof follows immediately from the ones of Theorems \ref{Th1}-\ref{Th3} by setting $\bm{G}_I$ and $\bm{F}_I$ to be null matrices, implying $\bm{P}_{\rangeperp{\bm{G}_I}}=\bm{P}_{\rangeperp{\bm{F}_I}}=\bm{I}_m.$ \IEEEQEDhere

A few comments on Theorem \ref{Th4} are in place here, most of which are analogous to those made in Remark \ref{VIPR1} below Theorems \ref{Th1}-\ref{Th3}:
\begin{remark} \label{VIPR2}
\begin{itemize}
\item The form (\ref{sl_free}) has been obtained and analyzed in \cite{Piotrowski2009}, \cite{Piotrowski2014}. The forms (\ref{sl2_free}) and (\ref{sl3_free}) are newly proposed.
\item For no rank constraint, i.e., $r=l$, Theorem \ref{Th4} yields alternative forms of the LCMV filter (\ref{blueish}) as
\begin{equation} \label{WLCMV_new}
  \bm{W}_{LCMV}=\bm{G}^{\dagger}\bm{R}^{-1/2}=\bm{F}^{\dagger}\bm{N}^{-1/2}.
\end{equation}
\item In view of (\ref{WLCMV_new}), the filters in (\ref{sl_free}), (\ref{sl2_free}) and (\ref{sl3_free}) differ only in the subspace that the estimate of the LCMV filter is orthogonally projected onto. Namely, they are of the form:
\begin{equation} \label{elegant2}
\bm{W}_r^\star=\bm{P}_{\bm{L}_r^{(i)}}\bm{W}_{LCMV},\ i=1,2,3.
\end{equation}
\item Furthermore, from (\ref{WLCMV_new}) we may obtain alternative representations of matrices $\bm{L}^{(1)}$ and $\bm{L}^{(3)}$ as:
  \begin{equation}
\bm{L}^{(1)}=\bm{W}_{LCMV}\bm{R}\bm{W}_{LCMV}^t-2\bm{Q},
  \end{equation}
  \begin{equation}
\bm{L}^{(3)}=\bm{W}_{LCMV}\bm{N}\bm{W}_{LCMV}^t,
  \end{equation}
with $\bm{L}^{(2)}=\bm{L}^{(1)}+2\bm{Q}=\bm{W}_{LCMV}\bm{R}\bm{W}_{LCMV}^t$ and $\bm{L}^{(4)}=\bm{L}^{(3)}-\bm{Q}=\bm{W}_{LCMV}\bm{N}\bm{W}_{LCMV}^t-\bm{Q}.$    
\item It is seen from the previous two points that the proposed filters, if cast in the form (\ref{elegant2}), require computation of the LCMV filter along with the orthogonal projection onto a minor subspace of one of $\bm{L}^{(i)}\in\mathbb{R}^{l\times l}$ matrices for $i=1,2,3$, which complexity is $\mathcal{O}(l^3)$ using na\"ive implementation~\cite{Golub1996}. Thus, the computational effort needed to compute the proposed filters in the form (\ref{elegant2}) is of the same order as of the LCMV filter. A minor subspace tracking method \cite{Chen1998} has been employed in a preliminary study \cite{Piotrowski2012b} for an online implementation of filter in (\ref{sl_free}).
\item We note that the filter (\ref{sl_free}) is expressed in terms of the covariance matrix $\bm{Q}$ of signal to be reconstructed through $\bm{L}^{(1)}$, whereas the filters in (\ref{sl2_free}) and (\ref{sl3_free}) do not depend explicitly on $\bm{Q}.$
\item From \cite[Lemma 1]{Piotrowski2014} we have
\begin{equation} \label{L1_2014_free}
  \bm{Q}=(\bm{H}^t\bm{R}^{-1}\bm{H})^{-1}-(\bm{H}^t\bm{N}^{-1}\bm{H})^{-1}.
\end{equation}
Then, an estimate of $\bm{Q}$ can be obtained through (\ref{L1_2014_free}) by replacing $\bm{R}$ and $\bm{N}$ with their finite samples estimates.
\item The MSE in (\ref{J1_free}), (\ref{J2_free}) and (\ref{J3_free}) yield natural rank-selection method applicable to all forms of the proposed filter by selecting $r$ which minimizes them. It should be highlighted that (\ref{J1_free}) depends on $r$ only through eigenvalues of the symmetric matrix $\bm{L}^{(1)}$, while (\ref{J2_free}) depends on $r$ through eigenvalues of the positive semidefinite matrix $\bm{L}^{(2)}$ and $tr(\bm{P}_{\bm{L}_r^{(2)}}\bm{Q}).$ Similarly, (\ref{J3_free}) depends on $r$ through eigenvalues of the positive semidefinite matrix $\bm{L}^{(3)}$ and $tr(\bm{P}_{\bm{L}_r^{(3)}}\bm{Q}).$
\end{itemize}
\end{remark}

\section{Numerical Examples} \label{ne}
\subsection{Simulation Framework}
In order to facilitate reproducibility of our research we
provide (jointly with this paper) a comprehensive simulation
framework that allows for estimation of error of signal
reconstruction for a number of spatial filters applied to MEG or EEG
signals. It is available for download at
\url{https://github.com/IS-UMK/supFunSim.git}. This framework may be used
directly from a single \href{http://www.orgmode.org}{org-mode} file (\texttt{supFunSim.org}) or as a set of MATLAB scripts.

\subsection{Signals in Source Space}
%
Generation of time series in source space for bioelectrical activity of brains'
cortical and subcortical regions is conducted using separate
MVAR models for the activity of interest $\q$ 
(\texttt{sim\_sig\_SrcActiv}), the interfering activity $\q_I$
(\texttt{sim\_sig\_IntNoise}), and the biological background noise $\q_b$
(\texttt{sim\_sig\_BcgNoise}).
Additionally, we also add Gaussian uncorrelated noise $\n_m$ to the
time series in sensor space
to mimic measurement noise and all the remaining activity of the brain
(\texttt{sim\_sig\_MesNoise}).

Both $\q$ and $\q_b$ are simulated using independent, random and stable MVAR models of order 6. MVAR modeling and fitting was performed using MVARICA \cite{MVARICA2008} and ARfit \cite{ARfit2001} packages. During generation of each MVAR model for $\q$, the 
coefficient matrix was multiplied by a
mask matrix
that has
80\% of its off-diagonal elements equal to zero. All the remaining diagonal and off-diagonal masking coefficients are equal to one. Such procedure allows us to define specific directed dependencies between activity of sources of interest. These dependencies are measured with the PDC which operates on coefficients derived from the MVAR model fitted to the signal. We refer the reader to \cite{Baccala2001} for detailed description of the PDC.
Activity of each of the $\q_I$ sources is generated as negative of the
$\q$ with added Gaussian noise of the same power as the $\q$ signal itself. In this way, we get the $\q_I$ signal to be correlated with $\q$.

\subsection{Volume Conduction Model and Leadfields} \label{vcm}
In order to obtain measurements (i.e., time series in sensor space), we first use
FieldTrip (FT) toolbox
\cite{FieldTrip2011}
for generation of
volume conduction model (VCM)
and
leadfields.
VCM is prepared (\texttt{ft\_prepare\_headmodel}) using \texttt{DIPOLI} method
\cite{FieldTripDipoli1989} that is applied to three triangulated
surface meshes (\texttt{tess\_innerskull.mat}, \texttt{tess\_outerskull.mat} and
\texttt{tess\_head.mat}) representing the
outer surfaces of
brain,
skull, and scalp, and they correspond to those provided as sample/reference data in the Brainstorm (BS) toolbox \cite{Brainstorm2011}.

Prior to VCM generation, we transform
coordinates of vertices of each mesh
to match spatial orientation and length unit of
the EEG system used.
Additionally, for the results presented herein, we arbitrarily select:
\begin{itemize}
\item \emph{HydroCel Geodesic Sensor Net} utilizing 128 channels as EEG cap layout.
\item Regions of interest (ROIs) corresponding to specific cortex patches, and whose geometry is reconstructed and parcellated from the detailed cortical surface by means of the BS toolbox (\texttt{tess\_fs\_tcortex\_15000V.mat}). We selected ROIs by
their anatomical description in Destrieux and Desikan-Killiany atlases \cite{FreeSurferCortex2004,FreeSurferCortex2006}. Surface parcellation was prepared using freely available
FreeSurfer Software
Suite \cite{FreeSurfer1999} such that each ROI is comprised of a triangular mesh and fully characterized by its nodes, from which candidates for source position and orientation (orthogonal to the mesh) will be randomly drawn.
\item Both thalami modeled jointly as a single triangulated mesh containing node candidates
for a random selection of subcortical bioelectrical activity. Also here, the orientation of sources is chosen as orthogonal to the mesh surface.
\end{itemize}

\begin{figure}[t]
\vspace*{-0.7cm}\hspace*{-1.4cm}\includegraphics[scale=0.27]{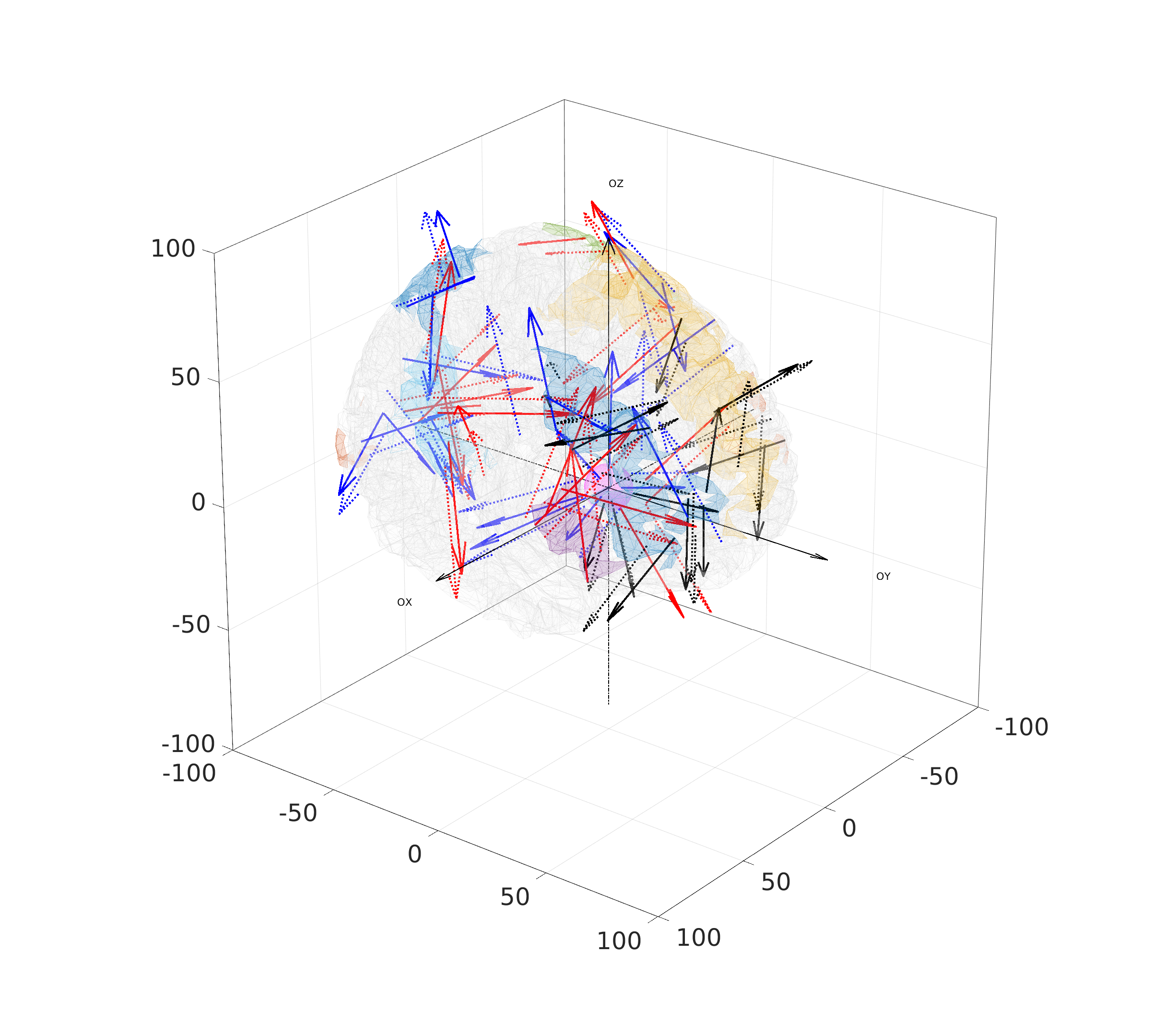}
  \caption{Example of simulation setup showing ROIs and vertices representing sources of interest $\q$ (black), interference $\q_I$ (red), and background activity $\q_b$ (blue). Depending on the simulation setup, ROIs and vertices can be fixed or randomly selected. Perturbation of source location and direction can also be introduced. The perturbed sources are indicated using dashed line.}
\end{figure}

\subsection{Signal in Sensor Space}
The forward model solution is based on the boundary element method (BEM) proposed in \cite{Fuchs2002}, which is implemented in the FT toolbox. Based on this solution, we obtain EEG measurements $\y$ using leadfields calculated for each source defined by its position and orientation. We note that for large power of $\q_I$, (\ref{model}) realistically represents signal $\y$, while the lower the power of $\q_I$, the more adequate (\ref{model_free}) becomes. 

\subsection{Simulations Settings and Configuration}
The simulations framework provided with this paper enables
to freely change essential simulation parameters.
These include, in particular:
\begin{itemize}
\item
  signal to interference noise ratio (SINR) defined as the ratio of power of $\q$ signal to the power of $\q_I$ signal, both projected onto sensor space,
\item
  signal to biological noise ratio (SBNR) defined as the ratio of power of $\q$ signal to the power of $\q_b$ signal, both projected onto sensor space,
\item
  signal to measurement noise ratio (SMNR) defined as the ratio of power of $\q$ signal projected onto sensor space to the power of $\n_m$ signal,
\item
  number of cortical ROIs to be randomly selected,
\item
  number of sources (of each type, per any of the cortical and subcortical regions described previously). For the sake of numerical results presented in this section, we selected $l=13$ sources of interest, $k=27$ sources of interfering activity, $p=27$ sources of background activity ($7$ on the cortex, and remaining $20$ in subcortical regions), 
\item
  number of simulation runs where a new MVAR model is generated in each run,
\item
  number of independent realizations based on each generated MVAR model (trials),
\item
  number of time samples per trial,
\item
  usage of original or perturbed leadfield matrices. In the latter case, leadfield vectors of sources of interest (columns of $\bm{H}$) and of interfering sources (columns of $\bm{H}_I$) may be generated such that:
  \begin{itemize}
  \item position of each source is shifted randomly within a cube of a given side length, centered at the original source location,
  \item orientation of each source is shifted randomly within a prescribed azimuth and elevation intervals, both centered at the original source orientation.
  \end{itemize}  
\end{itemize}
In the numerical example presented in this section, we considered patch-constraints, where the rank $s$ of $\bm{H}_{I_s}$ was set to $8$, which allowed us to keep approximately $30\%$ of singular values of $\bm{H}_I.$ These parameters are also adjustable in the proposed framework. Moreover, we also evaluated a perturbed leadfield matrix of sources of interest, where each source position was allowed to be shifted randomly within a cube of 20 mm side length, with azimuth and elevation of sources shifted by $\pi/32$ rad. The resulting perturbed leadfield matrix of sources of interest was denoted $\bm{H}_{PE}.$

For each combination of SNRs we conducted 1000 simulation runs. In each run, the location of the sources were
randomly chosen in accordance to the scheme described in Section \ref{vcm} and a new MVAR model was generated.
Each simulation run contained a realization of the MVAR process with 1000 samples, where:
\begin{itemize}
\item The first half (i.e., the first 500 samples) of each trial is interpreted as pre-task/stimulus
activity and is comprised of $\q_b$ signal projected onto sensor space along with $\n_m$ signal. The estimate of noise covariance matrix $\bm{N}$ is obtained from this section of the signals as a finite sample estimate.
\item The second half of each trial is comprised of all signals, i.e., $\q$, $\q_I$, $\q_b$ signals, projected onto sensor space along with $\n_m$ signal. The estimate of signal covariance matrix $\bm{R}$ is obtained from this second section of the signals as a finite sample estimate. 
\end{itemize}

\subsection{Performance Evaluation} \label{pe}
\subsubsection{Filters Evaluated} We selected for comparison the following spatial filters:
\begin{itemize}
\item The LCMV filter, expressed as both $$\bm{W}_{LCMV(\bm{R})}=(\bm{H}^t\bm{R}^{-1}\bm{H})^{-1}\bm{H}^t\bm{R}^{-1},$$ and $$\bm{W}_{LCMV(\bm{N})}=(\bm{H}^t\bm{N}^{-1}\bm{H})^{-1}\bm{H}^t\bm{N}^{-1},$$ 
\item The nulling filter in the form proposed in the work \cite{Hui2010}, namely $$\bm{W}_{NL}=[\bm{I}_l\ \bm{0}_{l\times k}](\bm{H}_c^t\bm{R}^{-1}\bm{H}_c)^{-1}\bm{H}_c^t\bm{R}^{-1},$$ \emph{cf.} (\ref{nullingish}).
\item The theoretically MSE-optimal MMSE (Wiener) filter, defined as $$\bm{W}_{F-MMSE}=\bm{Q}\bm{H}^t\bm{R}^{-1},$$ for the interference-free model, and $$\bm{W}_{I-MMSE}=\me{\q\q_c^t}\bm{H}_c^t\bm{R}^{-1},$$ for the model in presence of interference,
\item The zero-forcing filter, defined as $$\bm{W}_{ZF}=\bm{H}^\dagger,$$
\item The eigenspace-LCMV filters \cite{Sekihara2008} exploiting projection of the signal covariance matrix $\bm{R}$ onto its principal subspace of the forms $$\bm{W}_{EIG-LCMV(\bm{R})}=\bm{W}_{LCMV(\bm{R})}\bm{P}_{\bm{R}_{sig}},$$ and $$\bm{W}_{EIG-LCMV(\bm{N})}=\bm{W}_{LCMV(\bm{N})}\bm{P}_{\bm{R}_{sig}},$$
  where $\bm{P}_{\bm{R}_{sig}}$ is the orthogonal projection matrix onto subspace spanned by eigenvectors corresponding to $\la_1\geq\dots\geq\la_{sig}$ - the $sig$ largest eigenvalues of $\bm{R}$, where $sig$ is the dimension of signal subspace. In practice, it is difficult to estimate signal subspace dimension \cite{Sekihara2008}. Here, we selected its optimal value, i.e., the value of $sig$ that matches the number of active sources.
\end{itemize}

\subsubsection{Performance Measures}
We considered the following performance measures:
\begin{itemize}
\item \texttt{MSE}, defined as the squared Euclidean distance between the signal of interest and its reconstruction. For better clarity, both signal and its reconstruction were normalized prior to distance measurement. For all 13 signals of interest, the result is averaged across 500 time samples and 1000 simulation runs. 
\item \texttt{PDC}, defined as a correlation coefficient between original profiles of absolute PDC values and those based on MVAR model fitted to the reconstructed signal. The result is averaged across 13 signals of interest (each comprised of 500 samples) and 1000 simulation runs. 
\end{itemize}
The \texttt{MSE} and \texttt{PDC} performance measures are evaluated for:
\begin{itemize}
\item filters using original leadfield matrix of sources of interest $\bm{H}$: \texttt{MSE}($\bm{H}$) and \texttt{PDC}($\bm{H}$),
\item filters using perturbed leadfield matrix of sources of interest $\bm{H}_{PE}$: \texttt{MSE}($\bm{H}_{PE}$) and \texttt{PDC}($\bm{H}_{PE}$).
\end{itemize}
The above performance measures, along with others not reported here, are implemented in section \texttt{tg\_s06\_Error\_evaluation} of the provided simulation framework.

\subsubsection{Numerical Results}
Due to large number of filters considered, we present numerical results in Tables I-VI on the next page. 
The results presented therein show that for SINR=0 dB, the nulling filters outperform filters designed for interference-free model, which reflects observed signal less accurately in such settings. On the other hand, for higher values of SINR, the interference-free model gradually becomes the preferred one, and filters designed for this model achieve comparable or better performance than their counterparts derived for the model in presence of interference.

Moreover, the \texttt{MSE} and \texttt{PDC} performance measures used in simulations demonstrate that fidelity in reconstruction performance is related to accuracy of reconstructed \texttt{PDC} profiles, but there is no direct linkage between them. Indeed, the eigenspace-LCMV filters produce reconstruction particularly ill-suited to this aim. On the other hand, despite the fact that the proposed filters use mean-square error as the cost function, they also perform well in terms of \texttt{PDC} measure, which encourages their use in investigating interactions among sources based on MVAR model such as PDC or DTF \cite{Kus2004}. As discussed in Section \ref{interactions}, this is due to efficient cancellation of interfering activity, which enables more accurate fitting of the MVAR model to the reconstructed activity of sources of interest. We refer interested readers to \cite{Hui2010}, where this particular application of nulling filters is discussed in more detail. 

Unsurprisingly, the performance of most of the evaluated filters is negatively affected by using the perturbed leadfield matrix of sources of interest $\bm{H}_{PE}$ in place of the original leadfield matrix $\bm{H}$, for both MSE and PDC measures. In particular, degradation in MSE performance of the proposed filters is similar as in other filters evaluated, which is in line with numerical results of \cite{Piotrowski2009, Piotrowski2014}, especially considering that highly noisy settings (SMNR=10 dB and SBNR=0 dB) are used to obtain results presented in Tables I-VI. In the case of PDC, the performance of the proposed filters is very similar for $\bm{H}$ or $\bm{H}_{PE}.$ This intriguing fact should enhance usability of the proposed filters in the area of investigating interactions among sources based on MVAR model.

We should also emphasize that the presented results are for illustrative purpose and we invite readers to test other settings.

\subsection{Event-Related Experiments}
In the simulation framework proposed in this section we did not
insist on a specific type of EEG/MEG measurements. In particular,
the proposed framework is applicable to all types of EEG/MEG
measurements, including event-related EEG/MEG experiments, where one
may be interested in reconstructing activity which is evoked
(phase-locked) or induced (non-phase locked) to the stimulus. In such
experiments, we usually have stationarity in the covariance, but not
in the mean \cite{Moiseev2011}, due to presence of both evoked and
induced part of the brain response to the stimuli
\cite{David2006}. Then, if one is interested in reconstruction of
evoked (phase-locked) response, we recommend to use the filter
minimizing directly the power of reconstructed noise, i.e., the filter defined in (\ref{sl3}), if presence of interference is assumed. If the interference-free model is considered, then we recommend the filter in (\ref{sl3_free}). On
the other hand, if reconstruction of induced
(non-phase-locked) response is of interest, one may simply subtract the phase-locked
activity beforehand (see, e.g., \cite{Hui2010}).
\vspace{\baselineskip}

\section{Conclusions and Future Work}
We proposed two families of novel spatial filters parameterized by filter rank. We also provided a method to select this parameter such that the MSE of the filter output is minimized. The proposed filters extend the MV-PURE framework \cite{Yamada2006, Piotrowski2008, Piotrowski2009, Yamagishi2013, Piotrowski2014}. The approach of the proposed filters may be extended further with sensor space transformations, such as projecting $\bm{R}$ onto its principal subspace \cite{Gutierrez2006,Sekihara2008}.

We emphasize that the proposed filters can be applied in other areas of array signal processing such as wireless communications \cite{Piotrowski2009}. Indeed, such applications may stimulate further development of these filters, by deriving their online implementations or relaxing the nulling constraints (see, e.g., \cite{Piotrowski2012b, Yukawa2013} for related work).

\begin{onepage}
  \begin{center}
    \small
  \begin{tabular}{ | l? l | l |}
  \hline
Filter & \texttt{MSE}$(\bm{H})$ & \texttt{MSE}$(\bm{H}_{PE})$ \\ \Xhline{2pt}
LCMV\_R        &    1.47$\pm$0.18   &   1.52$\pm$0.14   \\ \hline
LCMV\_N	       &    1.35$\pm$0.25   &	1.34$\pm$0.24	\\ \hline
NL	       &    1.45$\pm$0.20   &	1.51$\pm$0.15	\\ \hline
MMSE	       &    1.38$\pm$0.19   &	1.45$\pm$0.15	\\ \hline
MMSE\_INT      &    \textbf{1.21$\pm$0.23}   &	1.34$\pm$0.18	\\ \hline
ZF	       &    1.53$\pm$0.20   &	1.51$\pm$0.20	\\ \hline
EIG\_LCMV\_R     &    1.28$\pm$0.22   &	\textbf{1.31$\pm$0.19}	\\ \hline
EIG\_LCMV\_N     &    1.37$\pm$0.26   &	1.35$\pm$0.24	\\ \Xhline{2pt}
MVP\_MSE        &    1.36$\pm$0.18   &	1.43$\pm$0.14	\\ \hline
MVP\_R	       &    1.32$\pm$0.17   &	1.43$\pm$0.14	\\ \hline
MVP\_N	       &    \textbf{1.12$\pm$0.26}   &	\textbf{1.16$\pm$0.24}	\\ \hline
MVP\_NL\_MSE    &    \textbf{1.11$\pm$0.14}   &	\textbf{1.26$\pm$0.12}	\\ \hline
MVP\_NL\_R      &    \textbf{1.14$\pm$0.17}   &	\textbf{1.32$\pm$0.17}	\\ \hline
MVP\_NL\_N      &    \textcolor{red}{\textbf{0.86$\pm$0.15}}   & \textcolor{red}{\textbf{0.95$\pm$0.15}} \\ \hline
\end{tabular} 
\captionof{table}{Spatial filters MSE performance for SMNR=10 dB, SBNR=0 dB and \textbf{SINR=0 dB}. The best result is in bold red and top 5 results are in bold.}
\end{center}
  \vspace{0.4cm}
  
\begin{center}
  \small
  \begin{tabular}{ | l? l | l |} 
  \hline
Filter & \texttt{MSE}$(\bm{H})$ & \texttt{MSE}$(\bm{H}_{PE})$ \\ \Xhline{2pt}
LCMV\_R        &   1.34$\pm$0.21   &   1.42$\pm$0.16   \\ \hline
LCMV\_N	       &   1.18$\pm$0.28   &   1.19$\pm$0.25   \\ \hline
NL	       &   1.39$\pm$0.21   &   1.47$\pm$0.16   \\ \hline
MMSE	       &   1.19$\pm$0.21   &   1.31$\pm$0.16   \\ \hline
MMSE\_INT      &   1.10$\pm$0.25   &   1.27$\pm$0.19   \\ \hline
ZF	       &   1.44$\pm$0.22   &   1.42$\pm$0.21   \\ \hline
EIG\_LCMV\_R   &   \textbf{1.08$\pm$0.25}   &   \textbf{1.15$\pm$0.21}   \\ \hline
EIG\_LCMV\_N   &   1.12$\pm$0.28   &   \textbf{1.13$\pm$0.25}   \\ \Xhline{2pt}
MVP\_MSE       &   1.13$\pm$0.17   &   1.27$\pm$0.14   \\ \hline
MVP\_R	       &   1.10$\pm$0.15   &   1.26$\pm$0.14   \\ \hline
MVP\_N	       &   \textcolor{red}{\textbf{0.80$\pm$0.21}}   &   \textcolor{red}{\textbf{0.88$\pm$0.20}}   \\ \hline
MVP\_NL\_MSE   &   \textbf{0.98$\pm$0.12}   &   \textbf{1.18$\pm$0.11}   \\ \hline
MVP\_NL\_R     &   \textbf{1.07$\pm$0.19}   &   1.27$\pm$0.19   \\ \hline
MVP\_NL\_N     &   \textbf{0.81$\pm$0.14}   &   \textbf{0.90$\pm$0.14}   \\ \hline
\end{tabular} 
\captionof{table}{Spatial filters MSE performance for SMNR=10 dB, SBNR=0 dB and \textbf{SINR=5 dB}. The best result is in bold red and top 5 results are in bold.}
\end{center}
\vspace{0.4cm}

\begin{center}
  \small
  \begin{tabular}{ | l? l | l |} 
  \hline
Filter & \texttt{MSE}$(\bm{H})$ & \texttt{MSE}$(\bm{H}_{PE})$ \\ \Xhline{2pt}
LCMV\_R        &   1.22$\pm$0.24   &   1.34$\pm$0.17   \\ \hline         
LCMV\_N	       &   1.07$\pm$0.29   &   \textbf{1.09$\pm$0.26}   \\ \hline
NL	       &   1.36$\pm$0.22   &   1.45$\pm$0.17   \\ \hline
MMSE	       &   1.02$\pm$0.22   &   1.21$\pm$0.17   \\ \hline
MMSE\_INT      &   1.04$\pm$0.26   &   1.24$\pm$0.19   \\ \hline
ZF	       &   1.39$\pm$0.23   &   1.38$\pm$0.21   \\ \hline
EIG\_LCMV\_R   &   \textbf{0.92$\pm$0.26}   &   \textbf{1.04$\pm$0.21}   \\ \hline
EIG\_LCMV\_N   &   \textbf{0.93$\pm$0.28}   &   \textbf{0.98$\pm$0.24}   \\ \Xhline{2pt}
MVP\_MSE       &   \textbf{0.93$\pm$0.15}   &   1.13$\pm$0.13   \\ \hline
MVP\_R	       &   0.94$\pm$0.14   &   1.15$\pm$0.13   \\ \hline
MVP\_N	       &   \textcolor{red}{\textbf{0.63$\pm$0.15}}   &   \textcolor{red}{\textbf{0.72$\pm$0.15}}   \\ \hline
MVP\_NL\_MSE   &   \textbf{0.93$\pm$0.12}   &   1.14$\pm$0.11   \\ \hline
MVP\_NL\_R     &   1.04$\pm$0.19   &   1.26$\pm$0.20   \\ \hline
MVP\_NL\_N     &   \textbf{0.80$\pm$0.14}   &   \textbf{0.89$\pm$0.14}   \\ \hline
  \end{tabular} 
\captionof{table}{Spatial filters MSE performance for SMNR=10 dB, SBNR=0 dB and \textbf{SINR=10 dB}. The best result is in bold red and top 5 results are in bold.}
\end{center}

\begin{center}
  \small
  \begin{tabular}{ | l? l | l |} 
  \hline
Filter & \texttt{PDC}$(\bm{H})$ & \texttt{PDC}$(\bm{H}_{PE})$ \\ \Xhline{2pt}
LCMV\_R        &    0.73$\pm$0.08   &   0.75$\pm$0.05   \\ \hline
LCMV\_N	       &    0.65$\pm$0.11   &   0.68$\pm$0.09	\\ \hline
NL	       &    0.72$\pm$0.09   &   0.74$\pm$0.06	\\ \hline
MMSE	       &    0.58$\pm$0.20   &   0.53$\pm$0.20	\\ \hline
MMSE\_INT      &    0.74$\pm$0.08   &   0.71$\pm$0.07	\\ \hline
ZF	       &    0.55$\pm$0.13   &   0.58$\pm$0.11	\\ \hline
EIG\_LCMV\_R     &    0.20$\pm$0.15   &   0.18$\pm$0.14	\\ \hline
EIG\_LCMV\_N     &    0.26$\pm$0.17   &   0.25$\pm$0.16	\\ \Xhline{2pt}
MVP\_MSE        &    \textbf{0.80$\pm$0.04}   &   \textbf{0.81$\pm$0.03}	\\ \hline
MVP\_R	       &    \textbf{0.79$\pm$0.06}   &   \textbf{0.79$\pm$0.05}	\\ \hline
MVP\_N	       &    0.75$\pm$0.09   &   0.75$\pm$0.07	\\ \hline
MVP\_NL\_MSE    &    \textbf{0.83$\pm$0.03}   &   \textcolor{red}{\textbf{0.83$\pm$0.02}}	\\ \hline
MVP\_NL\_R      &    \textcolor{red}{\textbf{0.84$\pm$0.03}}   &   \textcolor{red}{\textbf{0.83$\pm$0.03}}	\\ \hline
MVP\_NL\_N      &    \textcolor{red}{\textbf{0.84$\pm$0.03}}   &   \textcolor{red}{\textbf{0.83$\pm$0.03}}	\\ \hline
\end{tabular} 
\captionof{table}{Spatial filters PDC performance for SMNR=10 dB, SBNR=0 dB and \textbf{SINR=0 dB}. The best result is in bold red and top 5 results are in bold.}
\end{center}
\vspace{0.4cm}

\begin{center}
  \small
  \begin{tabular}{ | l? l | l |} 
  \hline
Filter & \texttt{PDC}$(\bm{H})$ & \texttt{PDC}$(\bm{H}_{PE})$ \\ \Xhline{2pt}
LCMV\_R        &   0.76$\pm$0.08   &   0.77$\pm$0.05   \\ \hline
LCMV\_N	       &   0.72$\pm$0.10   &   0.74$\pm$0.08   \\ \hline
NL	       &   0.73$\pm$0.09   &   0.75$\pm$0.06   \\ \hline
MMSE	       &   0.60$\pm$0.20   &   0.53$\pm$0.20   \\ \hline
MMSE\_INT      &   0.77$\pm$0.08   &   0.72$\pm$0.06   \\ \hline
ZF	       &   0.60$\pm$0.12   &   0.63$\pm$0.10   \\ \hline
EIG\_LCMV\_R   &   0.27$\pm$0.18   &   0.23$\pm$0.17   \\ \hline
EIG\_LCMV\_N   &   0.35$\pm$0.19   &   0.32$\pm$0.18   \\ \Xhline{2pt}
MVP\_MSE       &   \textbf{0.83$\pm$0.03}   &   \textbf{0.82$\pm$0.03}   \\ \hline
MVP\_R	       &   \textbf{0.83$\pm$0.04}   &   \textbf{0.82$\pm$0.03}   \\ \hline
MVP\_N	       &   \textbf{0.83$\pm$0.05}   &   \textbf{0.82$\pm$0.04}   \\ \hline
MVP\_NL\_MSE   &   \textcolor{red}{\textbf{0.84$\pm$0.02}}   &   \textcolor{red}{\textbf{0.83$\pm$0.02}}   \\ \hline
MVP\_NL\_R     &   \textcolor{red}{\textbf{0.84$\pm$0.03}}   &   \textcolor{red}{\textbf{0.83$\pm$0.03}}   \\ \hline
MVP\_NL\_N     &   \textcolor{red}{\textbf{0.84$\pm$0.03}}   &   \textcolor{red}{\textbf{0.83$\pm$0.03}}   \\ \hline
\end{tabular} 
\captionof{table}{Spatial filters PDC performance for SMNR=10 dB, SBNR=0 dB and \textbf{SINR=5 dB}. The best result is in bold red and top 5 results are in bold.}
\end{center}
\vspace{0.4cm}

\begin{center}
  \small
  \begin{tabular}{ | l? l | l |} 
  \hline
Filter & \texttt{PDC}$(\bm{H})$ & \texttt{PDC}$(\bm{H}_{PE})$ \\ \Xhline{2pt}
LCMV\_R        &   0.78$\pm$0.07   &   0.78$\pm$0.05   \\ \hline         
LCMV\_N	       &   0.76$\pm$0.09   &   0.77$\pm$0.07   \\ \hline
NL	       &   0.74$\pm$0.09   &   0.75$\pm$0.06   \\ \hline
MMSE	       &   0.61$\pm$0.20   &   0.51$\pm$0.22   \\ \hline
MMSE\_INT      &   0.78$\pm$0.07   &   0.73$\pm$0.06   \\ \hline
ZF	       &   0.63$\pm$0.12   &   0.65$\pm$0.10   \\ \hline
EIG\_LCMV\_R   &   0.32$\pm$0.20   &   0.27$\pm$0.19   \\ \hline
EIG\_LCMV\_N   &   0.42$\pm$0.21   &   0.35$\pm$0.19   \\ \Xhline{2pt}
MVP\_MSE       &   \textbf{0.84$\pm$0.03}   &   \textbf{0.83$\pm$0.02}   \\ \hline
MVP\_R	       &   \textbf{0.85$\pm$0.03}   &   \textbf{0.83$\pm$0.03}   \\ \hline
MVP\_N	       &   \textcolor{red}{\textbf{0.86$\pm$0.03}}   &   \textcolor{red}{\textbf{0.84$\pm$0.03}}   \\ \hline
MVP\_NL\_MSE   &   \textbf{0.84$\pm$0.02}   &   \textbf{0.83$\pm$0.02}   \\ \hline
MVP\_NL\_R     &   \textbf{0.84$\pm$0.03}   &   \textbf{0.83$\pm$0.03}   \\ \hline
MVP\_NL\_N     &   \textbf{0.84$\pm$0.03}   &   \textbf{0.83$\pm$0.03}   \\ \hline
  \end{tabular} 
\captionof{table}{Spatial filters PDC performance for SMNR=10 dB, SBNR=0 dB and \textbf{SINR=10 dB}. The best result is in bold red and top 5 results are in bold.}
\end{center}
\end{onepage}

We also developed open-source MATLAB simulation platform for evaluation of efficiency of spatial filters in reconstruction of brain activity from EEG/MEG measurements. This framework may be used
directly from a single \href{http://www.orgmode.org}{org-mode} file (\texttt{supFunSim.org}) or as a set of MATLAB scripts. \href{http://www.jupyter.org}{Jupyter} version is also planned and a detailed description of the framework will be reported in a future publication. It is available online at \url{https://github.com/IS-UMK/supFunSim.git} and may be extended according to users' needs. For example, various forms of regularization may be used to obtain better conditioning of $\bm{R}$ or $\bm{N}$, especially for small-sample size (see, e.g.,~\cite{Ledoit2004}). Such improvement over the currently implemented finite sample estimate of the covariance matrices can then be used across all implemented filters.

\appendices
\section{Known Results Used} \label{kru}
For convenience, let us recall that in this paper we assume that all eigen- and singular value decompositions considered have their eigen- and singular values organized in nonincreasing order.
\begin{fact}[$\mbox{Eckart-Young-Schmidt-Mirsky Theorem \cite{Mirsky1960}}$] \label{sat} 
  Let $\bm{A}\in\mn$ be a given matrix of $rk(\bm{A})=a$ and let us set rank constraint $r<a.$ Then, one has:
\begin{equation}  
\bm{X}_r^S\in\bigcap_{\uimal\in\ui}\left\{\arg\min_{\bm{X}_r\in\xr}\p \bm{X}_r-\bm{A}\p_\uimal\right\},
\end{equation}
if $\bm{X}_r^S$ is of the following form:
\begin{equation} \label{ewq3}
\bm{X}_r=[\bm{U}_{\bm{A}}]_{m\times r}[\bm{\Si}_{\bm{A}}]_{r\times r}([\bm{V}_{\bm{A}}]_{n\times r})^t,
\end{equation}
where $\bm{A}=\bm{U}_{\bm{A}}\bm{\Si}_{\bm{A}}\bm{V}_{\bm{A}}^t$ is a singular value decomposition of $\bm{A}.$ Further, $\bm{X}_r^S$ is a minimizer if and only if it is obtained in this way.
\end{fact}

\begin{fact}[$\mbox{\cite[p.113]{BenIsrael2003}}$] \label{BI2017} 
Let $\bm{A}\in\mn$ be a given matrix and let the subspace $S\subset\sgens{n}$ be given. Then, for any $b\in\sgens{m}$, $\bm{X}b$ is the minimum-norm least-squares solution of $\bm{A}x=b$, where $x\in S$, if and only if $\bm{X}=\bm{P}_s(\bm{A}\bm{P}_s)^\dagger$, where $\bm{P}_s$ is the orthogonal projection matrix onto $S.$ 
\end{fact}

\begin{fact}[\cite{Theobald1975}] \label{Theobald}
Let $\bm{C}\in\nn,\bm{D}\in\nn$ be symmetric matrices. Denoting by $c_1\geq c_2\geq\dots\geq c_n$ and $d_1\geq d_2\geq\dots\geq d_n$ the eigenvalues of $\bm{C}$ and $\bm{D}$, respectively, one has 
\begin{equation} \label{poeq}
tr(\bm{C}\bm{D})\leq\sum_{i=1}^n c_id_i.
\end{equation}
The equality can be attained.\footnote{The work \cite{Theobald1975} gives explicit form of eigenvalue decompositions of $\bm{C}$ and $\bm{D}$ for which equality can be attained.}
\end{fact}

\begin{remark} 
With notation as in Fact \ref{kru}-\ref{Theobald}, we have equivalently:
\begin{equation} \label{RkTheobald}
tr\big(\bm{C}(-\bm{D})\big)=-tr(\bm{C}\bm{D})\geq-\sum_{i=1}^n c_id_i.
\end{equation}
\end{remark}

\section{Proof of Theorem \ref{Th1}} \label{pdTh1}
  \paragraph{Change of variables} Let us first consider $\bm{G}:=\bm{R}^{-1/2}\bm{H}$ and $\bm{G}_I:=\bm{R}^{-1/2}\bm{H}_I$ along with $\bm{G}_c:=[\bm{G}\ \bm{G}_I]$ and $\bm{Z}_r:=\bm{W}_r\bm{R}^{1/2}.$ We note that $\bm{Z}_r\bm{G}=\bm{W}_r\bm{H}$ and $\bm{Z}_r\bm{G}_I=\bm{W}_r\bm{H}_I$ and consequently $\bm{Z}_r\bm{G}_c=[\bm{Z}_r\bm{G}\ \bm{Z}_r\bm{G}_I]=[\bm{W}_r\bm{H}\ \bm{W}_r\bm{H}_I]=\bm{W}_r\bm{H}_c.$ We also note that $rk(\bm{G})=l$, $rk(\bm{G}_I)=k$, $rk(\bm{G}_c)=l+k$ and $\bm{Z}_r\in\xrr{l}{m}$, as multiplication by invertible matrix does not change rank of matrix \cite{Horn1985}. Therefore, the optimization problem (\ref{rr_nulling}) can be recast in view of (\ref{mse}) and (\ref{argmin}) in terms of these new variables equivalently as:
\begin{equation} \label{rr_nulling_equiv}
  \left\{
  \begin{array}{ll}
    \zt{minimize} & \p \bm{Z}_r\p_F^2-2tr\big(\bm{Z}_r\bm{G}_c\me{\q_c\q^t}\big)\\
    \zt{subject to} & \left\{
    \begin{array}{l}
      \bm{Z}_r\in\ds{\bigcap_{\uimal\in\ui}}\arg\ds{\min_{\bm{Z}_r\in\xrr{l}{m}}}\p \bm{Z}_r\bm{G}-\bm{I}_l\p_\uimal^2\\
      \bm{Z}_r\bm{G}_I=\bm{0}_{l\times k}.\\
    \end{array}\right.\\
  \end{array}\right.
\end{equation}
In particular, we note that the constraint $\bm{Z}_r\bm{G}_I=\bm{0}_{l\times k}$ implies that
\begin{multline} \label{mvptrick}
\bm{Z}_r\bm{G}_c\me{\q_c\q^t}=\bm{Z}_r[\bm{G}\ \bm{G}_I]\me{\q_c\q^t}=\\ [\bm{Z}_r\bm{G}\ \bm{0}_{l\times k}]\me{\q_c\q^t}=\bm{Z}_r\bm{G}\bm{Q}, 
\end{multline}
which will be useful later on, with $\bm{Q}=\me{\q\q^t}.$
\paragraph{Feasible set} In this part we reformulate the definition of feasible set of $\bm{Z}_r$ satisfying constraints of optimization problem (\ref{rr_nulling_equiv}). Regarding the first constraint, we note that, from Fact \ref{sat} given in the Appendix \ref{kru}, we have
\begin{equation}
\bigcap_{\uimal\in\ui}\left\{\arg\min_{\bm{X}_r\in\xrr{l}{l}}\p \bm{X}_r-\bm{I}_l\p_\uimal\right\}=\bm{\La}_r\bm{\La}_r^t,
\end{equation}  
where $\bm{\La}_r=[\bm{\La}]_{l\times r}$ for an orthogonal matrix $\bm{\La}\in\sgen{l}{l}.$ We further note that equation $\bm{Z}_r\bm{G}=\bm{\La}_r\bm{\La}_r^t$ is solvable with respect to $\bm{Z}_r$, since, e.g., $\bm{\La}_r\bm{\La}_r^t\bm{G}^\dagger$ is a particular solution. Thus, $\bm{Z}_r$ satisfies $\bm{Z}_r\in\ds{\bigcap_{\uimal\in\ui}}\arg\ds{\min_{\bm{Z}_r\in\xrr{l}{m}}}\p \bm{Z}_r\bm{G}-\bm{I}_l\p_\uimal^2$ if and only if $\bm{Z}_r\bm{G}=\bm{\La}_r\bm{\La}_r^t$ for certain $\bm{\La}_r\in\sgen{l}{r}$ satisfying $\bm{\La}_r^t\bm{\La}_r=\bm{I}_r.$ Moreover, denoting $z^t_{r_i}$ to be the $i$-th column of $\bm{Z}_r^t$ for $i=1,\dots,l$, and upon noticing that the second constraint $\bm{Z}_r\bm{G}_I=\bm{0}_{l\times k}$ is equivalent to $z^t_{r_i}\in\nul{\bm{G}_I^t}=\rangeperp{\bm{G}_I}\subset\sgens{m}$ for $i=1,\dots,l$, it is seen that the constraints in (\ref{rr_nulling_equiv}) can be equivalently cast as
\begin{equation} \label{rr_nulling_constr}
  \left\{
    \begin{array}{l}
      \bm{Z}_r\bm{G}=\bm{\La}_r\bm{\La}_r^t\\
      z^t_{r_i}\in\rangeperp{\bm{G}_I}, i=1,\dots,l.\\
    \end{array}\right.\\
\end{equation}
In particular, from (\ref{mvptrick}) we obtain that it is sufficient to consider only minimum-norm solutions of (\ref{rr_nulling_constr}), as the cost function in (\ref{rr_nulling_equiv}) is, in view of (\ref{mvptrick}), such that
\begin{multline} \label{crucial_J}
  \p \bm{Z}_r^\star\p_F^2-2tr\big(\bm{Z}_r^\star \bm{G}\bm{Q}\big)=\\\p \bm{Z}_r^\star\p_F^2-2tr\big(\bm{\La}_r\bm{\La}_r^t\bm{Q}\big)\leq\\
  \p \bm{Z}_r^\diamond\p_F^2-2tr\big(\bm{\La}_r\bm{\La}_r^t\bm{Q}\big),
\end{multline}
where $\bm{Z}_r^\star$ is the minimum-norm solution of $\bm{Z}_r\bm{G}=\bm{\La}_r\bm{\La}_r^t$ and $\bm{Z}_r^\diamond$ is any other solution of $\bm{Z}_r\bm{G}=\bm{\La}_r\bm{\La}_r^t$, both subject to the constraints in (\ref{rr_nulling_constr}). 
\paragraph{Minimum-norm solution} We use now Fact \ref{BI2017} in the Appendix \ref{kru} and, after transposition, obtain that
\begin{equation} \label{arctic_star}
  \bm{Z}_r^\star=\bm{\La}_r\bm{\La}_r^t(\bm{P}_{\rangeperp{\bm{G}_I}}\bm{G})^\dagger \bm{P}_{\rangeperp{\bm{G}_I}}
\end{equation}  
is the least-squares solution of (\ref{rr_nulling_constr}) such that $\p \bm{Z}_r^\star\p_F^2$ is the smallest among all least-squares solutions of (\ref{rr_nulling_constr}). We proceed now to show that $\bm{Z}_r^\star$ is the minimum-norm solution of (\ref{rr_nulling_constr}). To this end, we note first that the second constraint in (\ref{rr_nulling_constr}) is satisfied by rows of $\bm{Z}_r^\star$ (columns of $(\bm{Z}_r^\star)^t$). Thus, it suffices to show that $(\bm{P}_{\rangeperp{\bm{G}_I}}\bm{G})$ is of full-column rank~$l$, which will ensure that $\bm{Z}_r^\star \bm{G}=\bm{\La}_r\bm{\La}_r^t$, as in such a case $(\bm{P}_{\rangeperp{\bm{G}_I}}\bm{G})^\dagger(\bm{P}_{\rangeperp{\bm{G}_I}}\bm{G})=\bm{I}_l$ \cite{Horn1985}.

To this end, we note that due to linear independence of columns of $\bm{G}_c=[\bm{G}\ \bm{G}_I]$, one has in particular that $\range{\bm{G}}\cap\range{\bm{G}_I}=\{0\}.$ Based on this fact, we will prove now that
\begin{equation} \label{littlewonder}
\zt{if }x\in\range{\bm{G}}\zt{ then }\bm{P}_{\rangeperp{\bm{G}_I}}x=0\iff\ x=0.
\end{equation}  
\non $(\Rightarrow)$ Let us express $x\in\range{\bm{G}}$ as $x=\bm{P}_{\range{\bm{G}_I}}x+\bm{P}_{\rangeperp{\bm{G}_I}}x.$ Then, by our assumption we have $x=\bm{P}_{\range{\bm{G}_I}}x$, hence $x\in\range{\bm{G}_I}.$ Thus, $x\in\range{\bm{G}}\cap\range{\bm{G}_I}$ and, from the fact that $\range{\bm{G}}\cap\range{\bm{G}_I}=\{0\}$, we conclude that it must be $x=0.$
\non $(\Leftarrow)$ This is obvious.

Based on (\ref{littlewonder}), it is easy to prove that $\bm{P}_{\rangeperp{\bm{G}_I}}$ is injective on $\range{\bm{G}}$ as follows: suppose that $x_1\in\range{\bm{G}},x_2\in\range{\bm{G}}$, $x_1\neq x_2$, and that $\bm{P}_{\rangeperp{\bm{G}_I}}x_1=\bm{P}_{\rangeperp{\bm{G}_I}}x_2.$ Thus, $\bm{P}_{\rangeperp{\bm{G}_I}}(x_1-x_2)=0$, hence from (\ref{littlewonder}) it must be $x_1=x_2$, which contradicts our assumption $x_1\neq x_2.$

Consider now the linearly independent set $\{g_1,g_2,\dots,g_l\}$ of columns of $\bm{G}.$ Since $\bm{P}_{\rangeperp{\bm{G}_I}}$ is injective on $\range{\bm{G}}$, the set $\{\bm{P}_{\rangeperp{\bm{G}_I}}g_1,\bm{P}_{\rangeperp{\bm{G}_I}}g_2,\dots,\bm{P}_{\rangeperp{\bm{G}_I}}g_l\}$ of columns of $(\bm{P}_{\rangeperp{\bm{G}_I}}\bm{G})$ is also linearly independent. This implies that $rk(\bm{P}_{\rangeperp{\bm{G}_I}}\bm{G})=l$, and consequently, that $(\bm{P}_{\rangeperp{\bm{G}_I}}\bm{G})^\dagger(\bm{P}_{\rangeperp{\bm{G}_I}}\bm{G})=\bm{I}_l.$ Thus, $\bm{Z}_r^\star$ in (\ref{arctic_star}) is the (minimum-norm) solution of $\bm{Z}_r\bm{G}=\bm{\La}_r\bm{\La}_r^t$ such that $z^t_{r_i}\in\rangeperp{\bm{G}_I}$ for $i=1,\dots,l.$ 
\paragraph{Selection of subspace} It follows from the previous considerations that we have narrowed the set of possible solutions of the optimization problem (\ref{rr_nulling_equiv}) to matrices of the form (\ref{arctic_star}), parameterized by
\begin{equation} \label{projection}
  \bm{P}_r:=\bm{\La}_r\bm{\La}_r^t,
\end{equation}
  which we recognize as an orthogonal projection matrix onto arbitrary subspace of $\sgens{l}$ of dimension $r.$ Therefore, we need to determine now the form of $\bm{P}_r$ minimizing the cost function in (\ref{rr_nulling_equiv}). Taking into account (\ref{mvptrick}) and using the fact that $\bm{P}_r$ and $\bm{P}_{\rangeperp{\bm{G}_I}}$ are both symmetric and idempotent \cite{Horn1985}, insertion of (\ref{arctic_star}) into cost function in (\ref{rr_nulling_equiv}) reveals the following:
\begin{multline} \label{revelation}
  \p \bm{Z}_r^\star\p_F^2-2tr\big(\bm{Z}_r^\star \bm{G}\bm{Q}\big)=\\
  tr\Big(\bm{P}_r(\bm{P}_{\rangeperp{\bm{G}_I}}\bm{G})^\dagger \bm{P}_{\rangeperp{\bm{G}_I}}\times\\\big(\bm{P}_r(\bm{P}_{\rangeperp{\bm{G}_I}}\bm{G})^\dagger \bm{P}_{\rangeperp{\bm{G}_I}}\big)^t\Big)-
  2tr(\bm{P}_r\bm{Q})=\\ tr(\bm{P}_r\bm{K}^{(1)}),
\end{multline}
where
\begin{equation} \label{K1_proof}
\bm{K}^{(1)}=(\bm{P}_{\rangeperp{\bm{G}_I}}\bm{G})^\dagger \bm{P}_{\rangeperp{\bm{G}_I}}\big((\bm{P}_{\rangeperp{\bm{G}_I}}\bm{G})^\dagger\big)^t-2\bm{Q}.
\end{equation}  
Clearly, $\bm{K}^{(1)}$ is a symmetric matrix, and so is $\bm{P}_r.$ Thus, by applying Fact \ref{Theobald} in the Appendix \ref{kru} along with (\ref{RkTheobald}) we obtain, upon setting $\bm{C}=\bm{P}_r$ and $\bm{D}=-\bm{K}^{(1)}$ therein that $tr(\bm{P}_r\bm{K}^{(1)})$ is minimized for $\bm{P}_r=\bm{P}_{\bm{K}_r^{(1)}}:=\bm{\La}_{\bm{K}_r^{(1)}}\bm{\La}_{\bm{K}_r^{(1)}}^t$, where $\bm{\La}_{\bm{K}_r^{(1)}}\in\sgen{l}{r}$ contains as its columns the eigenvectors corresponding to $\de_1^{(1)}\leq\dots\leq\de_r^{(1)}$ - the $r$ smallest eigenvalues of $\bm{K}^{(1)}.$ Then, the form of solution in (\ref{sl}) is obtained by inverting the change of variables to $\bm{W}_r^\star=\bm{Z}_r^\star \bm{R}^{-1/2}$, where $\bm{Z}_r^\star=\bm{P}_{\bm{K}_r^{(1)}}(\bm{P}_{\rangeperp{\bm{G}_I}}\bm{G})^\dagger \bm{P}_{\rangeperp{\bm{G}_I}}.$ Moreover, from (\ref{mse}), (\ref{rr_nulling_equiv}), (\ref{mvptrick}) and (\ref{revelation}) one has
\begin{multline} \label{mvpvalue}
\Jn_I(\bm{W}_r^\star)=\p \bm{Z}_r^\star\p_F^2-2tr\big(\bm{Z}_r^\star \bm{G}_c\me{\q_c\q^t}\big)+c=\\ tr(\bm{P}_{\bm{K}_r^{(1)}}\bm{K}^{(1)})+c=\sum_{i=1}^r\de_i^{(1)}+c,
\end{multline}
where $c=tr(\me{\q\q^t})=tr(\bm{Q}).$ \IEEEQEDhere

\section{Proof of Theorem \ref{Th2}} \label{pdTh2}
The proof proceeds analogously to the proof of Theorem~\ref{Th1}, where in place of optimization problem (\ref{rr_nulling_equiv}) we consider now, in view of (\ref{rr_nulling_R}), the following:
\begin{equation} \label{rr_nulling_equiv2}
  \left\{
  \begin{array}{ll}
    \zt{minimize} & \p \bm{Z}_r\p_F^2\\
    \zt{subject to} & \left\{
    \begin{array}{l}
      \bm{Z}_r\in\ds{\bigcap_{\uimal\in\ui}}\arg\ds{\min_{\bm{Z}_r\in\xrr{l}{m}}}\p \bm{Z}_r\bm{G}-\bm{I}_l\p_\uimal^2\\
      \bm{Z}_r\bm{G}_I=\bm{0}_{l\times k}.\\
    \end{array}\right.\\
  \end{array}\right.
\end{equation}
Then, for $\bm{Z}_r^\star$ of the form (\ref{arctic_star}), with $\bm{P}_r:=\bm{\La}_r\bm{\La}_r^t$ as in (\ref{projection}), we have now that
\begin{multline} \label{revelation2}
  \p \bm{Z}_r^\star\p_F^2=\\
  tr\Big(\bm{P}_r(\bm{P}_{\rangeperp{\bm{G}_I}}\bm{G})^\dagger \bm{P}_{\rangeperp{\bm{G}_I}}\big(\bm{P}_r(\bm{P}_{\rangeperp{\bm{G}_I}}\bm{G})^\dagger \bm{P}_{\rangeperp{\bm{G}_I}}\big)^t\Big)=\\tr(\bm{P}_r\bm{K}^{(2)}),
\end{multline}
where
\begin{multline} \label{K2}
\bm{K}^{(2)}=(\bm{P}_{\rangeperp{\bm{G}_I}}\bm{G})^\dagger \bm{P}_{\rangeperp{\bm{G}_I}}\big((\bm{P}_{\rangeperp{\bm{G}_I}}\bm{G})^\dagger\big)^t=\\\bm{K}^{(1)}+2\bm{Q}.
\end{multline}  
Clearly, $\bm{K}^{(2)}$ is a symmetric matrix. Thus, in order to find $\bm{P}_r$ minimizing $tr(\bm{P}_r\bm{K}^{(2)})$, we apply Fact \ref{Theobald} in the Appendix \ref{kru} along with (\ref{RkTheobald}) in exactly the same way as was done below (\ref{K1_proof}) in the proof of Theorem \ref{Th1}. Based on that, we obtain $\bm{P}_r=\bm{P}_{\bm{K}_r^{(2)}}:=\bm{\La}_{\bm{K}_r^{(2)}}\bm{\La}_{\bm{K}_r^{(2)}}^t$, where $\bm{\La}_{\bm{K}_r^{(2)}}\in\sgen{l}{r}$ contains as its columns the eigenvectors corresponding to $\de_1^{(2)}\leq\dots\leq\de_r^{(2)}$ - the $r$ smallest eigenvalues of $\bm{K}^{(2)}.$ Similarly as above, the form of solution in (\ref{sl2}) is obtained by inverting the change of variables to $\bm{W}_r^\star=\bm{Z}_r^\star \bm{R}^{-1/2}$, where $\bm{Z}_r^\star=\bm{P}_{\bm{K}_r^{(2)}}(\bm{P}_{\rangeperp{\bm{G}_I}}\bm{G})^\dagger \bm{P}_{\rangeperp{\bm{G}_I}}.$ Moreover, just as in (\ref{mvpvalue}), we obtain
\begin{multline} \label{mvpvalue2}
\Jn_I(\bm{W}_r^\star)=\p \bm{Z}_r^\star\p_F^2-2tr\big(\bm{Z}_r^\star \bm{G}_c\me{\q_c\q^t}\big)+c=\\ tr(\bm{P}_{\bm{K}_r^{(2)}}\bm{K}^{(1)})+c=tr\big(\bm{P}_{\bm{K}_r^{(2)}}(\bm{K}^{(2)}-2\bm{Q})\big)+c=\\\sum_{i=1}^r\de_i^{(2)}-2tr(\bm{P}_{\bm{K}_r^{(2)}}\bm{Q})+c,
\end{multline}
where $c=tr(\me{\q\q^t})=tr(\bm{Q}).$ \IEEEQEDhere

\section{Proof of Theorem \ref{Th3}} \label{pdTh3}
The closed algebraic form (\ref{sl3}) of the solution of the optimization problem (\ref{rr_nulling_N}) is obtained in exactly the same way as in the proof of Theorem \ref{Th2}, with $\bm{G}$ replaced by $\bm{F}$, $\bm{G}_I$ replaced by $\bm{F}_I$, and $\bm{G}_c$ replaced by $\bm{F}_c$, respectively. Thus, to complete the proof of Theorem \ref{Th3}, we only need to compute $\Jn_I(\bm{W}_r^\star)$, where $\bm{W}_r^\star=\bm{Z}_r^\star \bm{N}^{-1/2}$ with $\bm{Z}_r^\star=\bm{P}_{\bm{K}_r^{(3)}}\big(\bm{P}_{\rangeperp{\bm{F}_I}}\bm{F}\big)^{\dagger}\bm{P}_{\rangeperp{\bm{F}_I}}.$ Namely, using (\ref{mvptrick}) for $\bm{F}$ and $\bm{F}_I$, we have:
\begin{multline}
  \Jn_I(\bm{W}_r^\star)=tr(\bm{W}_r^\star \bm{R}(\bm{W}_r^\star)^t)-2tr\big(\bm{W}_r^\star \bm{H}_c\me{\q_c\q^t}\big)+c=\\
  tr(\bm{W}_r^\star\underbrace{(\bm{H}_c\bm{Q}_c\bm{H}_c^t+\bm{N})}_{\bm{R}}(\bm{W}_r^\star)^t)-2tr\big(\bm{W}_r^\star \bm{H}_c\me{\q_c\q^t}\big)+c=\\
  tr(\bm{Z}_r^\star \bm{F}_c\bm{Q}_c\bm{F}_c^t(\bm{Z}_r^\star)^t)+tr(\bm{Z}_r^\star(\bm{Z}_r^\star)^t)-2tr\big(\bm{Z}_r^\star \bm{F}_c\me{\q_c\q^t}\big)+c=\\
  tr(\bm{Z}_r^\star \bm{F}\bm{Q}\bm{F}^t(\bm{Z}_r^\star)^t)+tr(\bm{Z}_r^\star(\bm{Z}_r^\star)^t)-2tr\big(\bm{Z}_r^\star \bm{F}\bm{Q}\big)+c=\\
  tr\big(\bm{P}_{\bm{K}_r^{(3)}}\bm{Q}(\bm{P}_{\bm{K}_r^{(3)}})^t\big)+\p \bm{Z}_r^\star\p_F^2-2tr\big(\bm{P}_{\bm{K}_r^{(3)}}\bm{Q}\big)+c=\\
  \p \bm{Z}_r^\star\p_F^2-tr(\bm{P}_{\bm{K}_r^{(3)}}\bm{Q})+c=\\
  tr(\bm{P}_{\bm{K}_r^{(3)}}\bm{K}^{(4)})+c,
\end{multline}
where $c=tr(\me{\q\q^t})=tr(\bm{Q})$ and
\begin{multline}
  \bm{K}^{(4)}=(\bm{P}_{\rangeperp{\bm{F}_I}}\bm{F})^\dagger \bm{P}_{\rangeperp{\bm{F}_I}}\big((\bm{P}_{\rangeperp{\bm{F}_I}}\bm{F})^\dagger\big)^t-\bm{Q}=\\ \bm{K}^{(3)}-\bm{Q}.
\end{multline}
Finally,
\begin{multline}
\Jn_I(\bm{W}_r^\star)=tr\big(\bm{P}_{\bm{K}_r^{(3)}}(\bm{K}^{(3)}-\bm{Q})\big)+c=\\\sum_{i=1}^r\de_i^{(3)}-tr(\bm{P}_{\bm{K}_r^{(3)}}\bm{Q})+c.\ \IEEEQEDhere
\end{multline}  



\section*{Acknowledgment}
The authors are grateful to anonymous reviewers for their constructive comments which surely promoted the readability of the revised manuscript. 

\bibliographystyle{IEEEtran}
\bibliography{IEEEabrv,references2,loc-ref}

\vspace*{-5\baselineskip}

\end{document}